\documentclass{aa}  

\usepackage{graphicx}
\usepackage{txfonts}
\usepackage{lscape}
\usepackage{color}
\usepackage{hyperref}
\usepackage{tikz}
\usepackage{subcaption}
\newcommand{\orcit}[1]{\protect\href{https://orcid.org/#1}{\protect\includegraphics[width=8pt]{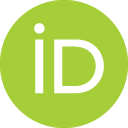}}}

\begin{document}

   \title{Stellar Population Astrophysics (SPA) with the TNG\thanks{Based on observations made with the Italian Telescopio Nazionale Galileo (TNG) operated on the island of La Palma by the Fundación Galileo Galilei of the INAF (Istituto Nazionale di Astrofisica) at the Spanish Observatorio del Roque de los Muchachos. This study is part of the Large Program titled SPA – Stellar Population Astrophysics: the detailed, age-resolved chemistry of the Milky Way disc (PI: L. Origlia), granted observing time with HARPS-N and GIANO-B echelle spectrographs at the TNG.}}

   \subtitle{$\alpha$-elements, lithium, sodium and aluminum in 16 open clusters}

  \author{R. Zhang
          \inst{1,2}
          \and
 S. Lucatello\inst{2,3}
 \and
 A. Bragaglia\orcit{0000-0002-0338-7883}\inst{4}
  \and
 J. Alonso-Santiago\inst{5}
 \and
 G. Andreuzzi\inst{6,7}
 \and
 G. Casali\inst{4,11}
 \and\
 R. Carrera\inst{2}
 \and
 E. Carretta\orcit{0000-0002-1569-1911}\inst{4}
  \and
 V. D'Orazi\inst{2}
\and
 A. Frasca\inst{5}
  \and
 X. Fu\orcit{0000-0002-6506-1985}\inst{9,4}
 \and
 L. Magrini\inst{10}
 \and
 I. Minchev\inst{12}
 \and
 L. Origlia\inst{4}
 \and
 L. Spina\inst{2}
 \and
 A. Vallenari\inst{2} 
}
   \institute{Dipartimento di Fisica e Astronomia, Universit\'a di Padova, vicolo Osservatorio 2, 35122, Padova, Italy \\
   \email{ruyuan.zhang@studenti.unipd.it}
   \and
   INAF-Osservatorio Astronomico di Padova, vicolo Osservatorio 5, 35122, Padova, Italy
   \and
   Institute for Advanced Studies, Technische Universit{\"a}t M{\"u}nchen, Lichtenbergstra{\ss}e 2 a, 85748 Garching bei M{\"u}nchen
   \and
   INAF-Osservatorio di Astrofisica e Scienza dello Spazio di Bologna, via P. Gobetti 93/3, 40129, Bologna, Italy
  \and
  INAF-Osservatorio Astrofisico di Catania, Via S. Sofia 78, 95123, Catania, Italy 
  \and           
   Fundaci\'on Galileo Galilei - INAF, Rambla Jos\'e Ana Fernández P\'erez 7, 38712, Breña Baja, Tenerife, Spain
  \and
  INAF-Osservatorio Astronomico di Roma, Via Frascati 33, 00078, Monte Porzio Catone, Italy
  \and
  Dipartimento di Fisica e Astronomia, Universit\'a degli Studi di Firenze, Via G. Sansone 1, 50019, Sesto Fiorentino, Firenze, Italy 
  \and
  KIAA-The Kavli Institute for Astronomy and Astrophysics at Peking University, Beijing 100871, China
  \and
  INAF-Osservatorio Astrofisico di Arcetri, Largo E. Fermi, 5, 50125, Firenze, Italy
  \and
  Dipartimento di Fisica e Astronomia, Universit\'a degli Studi di Bologna, Via Gobetti 93/2, I-40129 Bologna, Italy
  \and
  Leibniz Institute for Astrophysics Potsdam, An der Sternwarte 16, 14482 Potsdam, Germany
            }

   \date{Received ; accepted }

  \abstract
    {Exploring the Galactic chemical evolution and enrichment scenarios with open clusters (OCs) allows us to understand the history of the Milky Way disk. High-resolution spectra of OCs are a crucial tool, as they provide precise chemical information, to combine with precise distances and ages.}
   {The aim of the Stellar Population Astrophysics (SPA) project is to derive homogeneous and accurate comprehensive chemical characterization of a number of poorly studied OCs. }
  {Using the HARPS-N echelle spectrograph at the Telescopio Nazionale Galileo (TNG), we obtained high-resolution spectra of giant stars in 18 OCs, 16 of which are chemically characterized for the first time, and two of which are well studied for comparison. The OCs in this sample have ages from a few tens of Myr to 4 Gyr, with a prevalence of young clusters. We already presented the radial velocities and atmospheric parameters for them in a previous SPA paper. Here, we present results for the $\alpha$-elements O, Mg, Si, Ca and Ti, and the light elements Na and Al, all determined by the equivalent width method. We also measured Li abundance through the synthesis method. 
  }
{We discuss the behaviors of lithium, sodium and aluminum in the context of stellar evolution. For Na and Al, we compare our findings with models to investigate their behaviors as a function of mass, suggesting that Na 
mixing to the surface might start in masses as low as 2\,M$_{\sun}$.
We study the radial, vertical, and age trends for the measured abundance ratios in a sample that combines our results and recent literature for OCs, finding significant  (positive) gradients only for [Mg/Fe] and [Ca/Fe] in all cases. Finally, we compare O and Mg in the combined sample with chemo-dynamical models, finding a good agreement for intermediate-age and old clusters. There is a sharp increase in the abundance ratios measured among very young clusters (age $<$ 300 Myr ), accompanied by a poorer fit with the models for O and Mg, likely related to the inadequacy of traditional model atmospheres and methods in the derivation of atmospheric parameters and abundance ratios for stars of such young ages}
{}


   \maketitle
%
\section{Introduction}

 Open clusters  (OCs) are regarded as a simple stellar population, or in other words, their stars were born in the same episode of star formation, and thus they share similar properties such as age, distance and composition. This makes OCs ideal laboratories for exploring stellar evolution in different phases and for stars of different masses. 
This also means that the study of just a small subset of their members is sufficient in order to characterize the chemistry of an OC.  It is thus relatively easy to collect data on large samples of OCs, a fact that is exploited in the study of the Galactic disk. 

Open clusters are found in the Galactic thin disk, and, unlike field stars, their ages can be measured with good accuracy \citep[e.g.,][]{bt06,Bossini2019,cantat20}.
Open clusters are therefore useful for measuring the disk metallicity distribution at different distances from the Galactic center, in different azimuthal directions, and at different ages. This also applies to other chemical abundances; studying  species produced by different nucleosynthetic channels and by different progenitors, is important in order to understand the disk formation and chemical enrichment scenarios \citep{spina22, sharma21,vincenzo21,sun20,ishigaki12,magrini09,zhong20, weinberg19}.


The Gaia second data release (DR2) and the third early data release (EDR3) contain precise information about the position, parallax, proper motions, and photometry of more than 1.8 billion stars, including those in stellar clusters. The use of Gaia astrometry, possibly combined with auxiliary ground-based information on radial velocity, provides accurate membership information for OC stars \citep[see e.g.,][respectively]{cantat2018,jackson22}.
Presently, only a small fraction, about $~10\%$, of the OCs in the Milky Way have been observed by high-resolution spectroscopy, and with more objects being identified by Gaia\footnote{Hundreds of new OCs have been identified based on Gaia data, 
\citep[see e.g.,][]{castro2019,castro20,castro21,sim19,he21}. In addition, large scale structures have been discovered, \citep[see e.g.,][]{kounkel19}.},
further observations with very high-quality data are crucial in order to further investigate the structure of the thin disk.

Gaia also has spectroscopic capabilities \citep[see e.g.,][]{sartoretti18,cropper18}. However, those spectra, which are taken at relatively low resolution (R=11500) and in a narrow wavelength range near the infrared calcium triplet (845-872 nm), only enable the determination of a limited number of elements, with rather large uncertainties.
On the other hand, accurate composition information is necessary not only to study the chemical properties, but also to reliably measure the age of the OCs \citep[see e.g.,][]{Bossini2019}, with obvious repercussions.  

Large spectroscopic surveys such as Gaia-ESO, APOGEE, and GALAH 
 are providing data on OCs \citep[see e.g.,][]{randich22,donor20,spina21}.
However, those surveys generally have limited spectral coverage and/or insufficient spectral resolution to derive a full, accurate chemical characterization (a partial exception is Gaia-ESO, at least for the small fraction of stars observed with UVES-FLAMES at high resolution, R=47000).
Thus, important contributions can come from other, smaller-scale projects, dedicated to OCs. These projects provide the chance to collect spectra with high enough resolution and wide enough coverage to investigate the detailed chemical composition of OCs, probing all of the main nucleosynthetic channels. An example is the OCCASO project \citep{casamiquela19,carrera22}, where about 50 OCs have been studied using different spectrographs with R>50000. Another one is the One Star to Tag Them All (OSTTA) project, which uses FIES at NOT (R=65000) and FLAMES at VLT (R=45000), with more than 50 OCs observed \citep[see][for the first results]{carreraNOT}.

On the same line is our project, Stellar Populations Astrophysics (SPA), at the Italian Telescopio Nazionale Galileo (TNG), using the HARPS-N and GIANO echelle spectrographs.  
The present work is part of a series of SPA papers dedicated to OCs. The aim is to use a sample of OCs as a tool to investigate key properties of the Galactic thin disk.  The precision of the chemical abundances derived from the high-resolution spectra, combined with the ages and distances of OCs, allows the age-composition-distance relationships to be probed. In \cite{paperI} we derived the atmospheric parameters and Fe abundances of 16 poorly studied OCs. We probed the age-metallicity and distance-metallicity gradients, comparing our results to the latest chemo-dynamical models for the Galactic disk. In the present paper, we focus on $\alpha$-elements (O, Mg, Si, Ca and Ti\footnote{While Ti is not an $\alpha$-element, it is known to observationally behave as such, and is thus grouped together with Mg, Si, and Ca to measure an average $\alpha$-elements abundance.}),
on the light elements Na and Al, and on Li. 

Our choice to focus on these elements is due to the fact that $\alpha$-elements are crucial for probing the disk, as [$\alpha$/Fe] is used to identify the stars not only from the halo or Galactic disk, but also from the thin disk or thick disk.
Furthermore,  Li, Na, and Al provide insight into stellar evolution. On the one hand, Li is a very fragile element, very sensitive to details of the treatment of the mixing. On the other hand, Na and Al surface composition in giants of young OCs can, in fact, be different from their initial (main sequence) composition, depending on their masses \citep[i.e., on cluster ages, see e.g.,][]{ventura13,lagarde12}.




In Section 2 we describe 
the chemical abundance analysis procedure, together with a comparison between the present work and the literature. In Section 3  the chemical distribution and the comparison with the chemo-dynamical models are presented. Finally, a summary and conclusions are presented in Section 4.



\section{Derivation of chemical abundances}

All data for the SPA targets in this work were obtained from the 3.5\,m TNG in La Palma, with the high-resolution spectrograph HARPS-N (R=115000, wavelength coverage from 3830 to 6930 $\AA$). The observations were collected between December 2018 and December 2019. Data reduction was performed by the observatory pipeline, and spectral continuum normalization and combination were carried out with IRAF\footnote{A software released by the 
National Optical Astronomy Observatory (NOAO), which tool is used to reduce and analysis the astronomical data.}.
As described in \citet{paperI}, the stellar parameters were derived by equivalent width (EW) analysis measured with ARES \footnote{https://github.com/sousasag/ARES}  \citep{2015A&A...577A..67S}, using the  Local Thermodynamic 
 Equilibrium (LTE) code MOOG code \citep{sneden73} -- in its version pymoogi \footnote{https://github.com/madamow/pymoogi}, a python wrapper on MOOG 2019 --  combined with the ATLAS9 \citep{2003IAUS..210P.A20C} atmospheric grid.
For more details on target selection observations, data reduction, and stellar parameter derivation we refer the reader to \citet{paperI}.

Throughout the paper we use the standard spectroscopic notation, that is, for any given species X, [X] = $\log \epsilon(X)_{star} - \log \epsilon(X)_\odot$, [X/Fe] = [X/H] $-$[Fe/H], 
and $\log \epsilon(X) \equiv A(\epsilon) = \log (N_X/N_H) + 12.0$ for absolute number density abundances.

\subsection{Adopted solar abundance}
As solar reference abundance we adopt the default one in MOOG, (see e.g., \citet{asplund09}). We did, however, perform the analysis of a solar  spectrum, also obtained with HARPS-N, using the same steps as the SPA stars, deriving  the following values for the elements under discussion:
A(Li)=0.96$\pm$0.10,
A(O)=8.74$\pm$0.04,
A(Na)=6.18$\pm$0.08, 
A(Mg)=7.58$\pm$0.09, 
A(Al)=6.44$\pm$0.07, 
A(Si)=7.52$\pm$0.03, 
A(Ca)=6.26$\pm$0.05, 
and A(Ti)=4.89$\pm$0.06 (and A(Fe)=7.50, as in \citealt{paperI}). 

The values are generally in good agreement with those in \cite{asplund09}.
An exception is O, which is in marginal agreement (the discrepancy may be also due to the different modeling of Asplund, i.e.,  3D versus 1D). Its abundance was derived from the two forbidden lines for the SPA stars, which yield A(O)=8.715 from 6300.31 $\AA$ and A(O)=8.768 from 6363.79 $\AA$ for the Sun. It is worth noting that in fact, the 6300 A line is affected by blending with a Ni line and the 6363 $\AA$ by a CN  line. \cite{caffau08} also find a discrepancy between the two lines, using spectra of much higher resolution (R> 300,000), and estimate the blends to account for up to 37\% and 19\%  \citep{caffau13} of the lines, respectively. The O abundance in our solar abundance is the average of the abundances from the two lines.



\subsection{Lithium }
 In our spectra,  the Li~{\sc i} line near 6707.81 $\AA$ is generally weak and mildly affected by blending with Fe~{\sc i} and CN lines. Measurements for Li abundances were derived by the synthesis method using MOOG and the line list from \cite{dorazi15}.
 
 For five stars in our sample we could only place upper limits no the Li abundances, three of them belonging to the OC that cluster we have, NGC 2682; they are all, however, less evolved objects.
 Nonlocal thermodynamic 
 equilibrium (NLTE) corrections for lithium were estimated using the same tool as for Na (see Sec.~2.4). All the corrections are positive, with values ranging from 0.094 to 0.361 dex.  
 The uncertainties associated with the Li measurements are a combination of the synthesis fitting error and of the uncertainties of stellar parameters. Errors are listed in Table \ref{tab:Li}, while sensitivities to the parameters are listed in Table \ref{tab:sens2}. 

 \subsection{$\alpha$-elements}
 Abundances for the $\alpha$-elements O, Mg, Si, Ca, and Ti 
 are presented in this work. The line list used to derive the abundances is from 
 \cite{dorazi20}, and covers the range between 3940 $\AA$ and 6900 $\AA$.  Equivalent widths were measured with ARES \citep{2015A&A...577A..67S}.
  Manual measurements with IRAF were performed for strong lines (EW > 150 mA), lines with high fitting errors from ARES, and lines whose ARES EWs resulted in highly discrepant abundances. This check was done to account for the possible poor automatic fitting or for line blending. 

Particular care was required for Oxygen in line measurements.
The two oxygen lines (6300.31 $\AA$\, and 6363.79 $\AA$), used to perform the O abundance analysis in these stars, fall in a spectral range affected by telluric lines. 
Hence, depending on the radial velocity of the star, the O lines are affected by different degrees of blending. While modeling of telluric absorption is possible, in the case of blending the resulting O lines (and in turn O abundances) have considerable uncertainties even after telluric absorption  corrections have been applied.

However, the radial velocities for our sample excluded the contamination of the 6363 \AA\, line, while the 6300 \AA\, line was affected by telluric contamination in three cases.
Therefore, we chose  to discard the affected lines rather than subtract the modeled telluric absorption, and rely on the clean lines to derive the O abundances.
We measured O abundance from one line in three stars and from two lines in 37 stars.  We compared the oxygen abundance from 6300.31 $\AA$\, with that from 6363.79 $\AA$, and we found a statistically insignificant difference. \cite{caffau13} reports a discrepancy between the two lines in dwarfs, decreasing but still significant among warm giants, down to T$_{eff}\sim$5300 K, while the discrepancy seems to disappear below this effective temperature. Our results seem to be consistent with that.

We note that we have not taken into account the Ni or CN lines that are known to have a non-negligible effect on the solar O abundances derived from the two lines in consideration. We have estimated the effect of including those features in the analysis (assuming CN typical for clump stars and solar-scaled Ni) as, at most, a global increase of $\sim$0.1\,dex in O, generally within the error. However, we note that the oscillator strength associated with the Ni line could have considerable uncertainty (see e.g., \cite{caffau13}. 

Magnesium abundances are typically derived from five lines, for two of which (4703 and 5528) EWs were consistently measured manually. The lines were generally in good agreement with each other with no evidence for lines yielding systematically discrepant abundances. A similar number of lines, in excellent agreement with each
other, were used for Si.
Calcium is based on at least ten lines in all the stars in the sample, and at least 18 in half of the sample. Lines are in excellent agreement.

 Titanium {\sc i} and {\sc ii}  are based on a good number of transitions (12 to 42 and 12 to 30, respectively). The agreement between the species in the two ionized states is less than optimal, with Ti {\sc ii} resulting systematically enhanced with respect to Ti {\sc i}, with 
 A(Ti~{\sc i})-A(Ti~{\sc ii})=-0.32$\pm$0.18 dex. 

\cite{baratella2020} also find Ti~{\sc ii} overabundant  with respect to Ti~{\sc i} when using atmospheric parameters derived with the same traditional approach that we use (1D, LTE, Fe line based). In their paper, they argue that the traditional (1D, LTE analysis based on Fe lines) is not optimal for very young dwarfs and they use an approach based on Fe and Ti lines to derive the parameters, considerably lessening the disagreement between the neutral and ionized species of Ti, which both have result similar to the Ti~{\sc i} abundance derived with the traditional method.

We find a trend in the offset with age, with younger clusters having, on average, a worse match between the two ionization states of Ti. However, the offset also correlates with evolutionary stage gravity, suggesting that blends also play a role, given that low-gravity, cooler stars have more crowded spectra. The spectra of cooler, low-gravity stars are more crowded, leading to larger uncertainties on the measurement of ionized Ti, which is based on fewer and weaker lines than its neutral counterpart. Lastly, it is worth mentioning that NLTE corrections might also play a role in this.
For these reasons, we assume that the Ti~{\sc ii} is the less reliable of the two measurements, and thus we use only Ti~{\sc i} in further discussions. 

 
 \subsection{Sodium and aluminum}
For sodium abundance analysis, LTE is a poor assumption for line formation for most of the transitions used in the optical. Abundance ratios [Na/Fe] from optical spectra for the stars of the kind under discussion are generally overestimated in LTE, leading to different LTE and NLTE trends of chemical abundance with Galactocentric distance \citep[see e.g,][]{lind11}. We use the lines at 4751$\AA$, 5148$\AA$, 5682$\AA$, 5688$\AA$, 6154$\AA$, and 6160$\AA$, and for each star, derived the NLTE corrections using the tool \footnote{http://www.inspect-stars.com/} on a line-by-line basis, applied to the individual line abundances that were then averaged to determine the Na abundance for the star. All the corrections are negative and vary from $-$0.216 to $-$0.005 dex.

The Al abundance was calculated based on the doublet at 6696.03 and 6698.67 \AA. The NLTE corrections for Al were calculated according to the prescriptions in \cite{Nordlander2017}, using the code kindly provided by the authors. 
In general, the corrections are negative. 
The Al line at 6696.03~\AA\, has a range of NLTE corrections between $-0.03$ and $-0.09$ dex, while the corrections range from $-0.02$ to $-0.06$ dex for the 6698.67~\AA \, line.

\subsection{Uncertainties on measured abundance ratios}
Uncertainties associated with the abundances were derived using the same approach used in \citet{paperI}.
 In order to derive the sensitivity to the atmospheric parameters of the measured abundances, we selected nine stars as representative of the sample (one for each 100 K bin). The variations used were $\Delta$T$_{\textrm{eff}}$=200 K, $\Delta$log g=0.2 dex, $\Delta$[Fe/H]=0.2 dex, and $\Delta$ v$_{micro}$=0.4 km/s. The derived abundances for the $\alpha$-elements are listed in  Table \ref{tab:A1} and Table \ref{tab:A2}, while abundances for Li, Na, and Al, that is, elements requiring NLTE correction, are listed in Table~\ref{tab:Li}. The sensitivity to the parameters' uncertainties is listed in Table~\ref{tab:sens2} for Li, Na, and Al, and in Table~\ref{tab:sens1} for the $\alpha$-elements.

%

\subsection{Comparison with the literature}


The abundance of Li is known to change considerably throughout the evolution of the star, and therefore a comparison is meaningful only in the case of stars 
that have very similar atmospheric parameters.
There is very limited information in the literature on the Li content in the present OC sample (with the exception of NGC~2682). Just one other star has a  Li measurement previously reported in the literature in our entire sample: Collinder 350\_2, which star has log$\epsilon$(Li)$=$1.41$\pm$0.03 dex from \cite{casali20}, which is compatible within the error with our result of 1.58$\pm$0.14 dex. 
Out of the four NGC 2682 stars in our sample, we could measure Li only in one star,  with a value of 0.461$\pm$0.30 dex. We place upper limits on the other three. However, while the Li content of NGC~2682 has been extensively studied in the literature \citep[see e.g.,][]{Pace2012,Marilia2020,magrini21a}, 
there are no reported Li measurements for any of these stars, nor for stars of very similar atmospheric parameters. Therefore, a direct comparison is not meaningful.

\begin{figure*}[htbp]
   \centering
   \includegraphics[height=18cm,width=18cm]{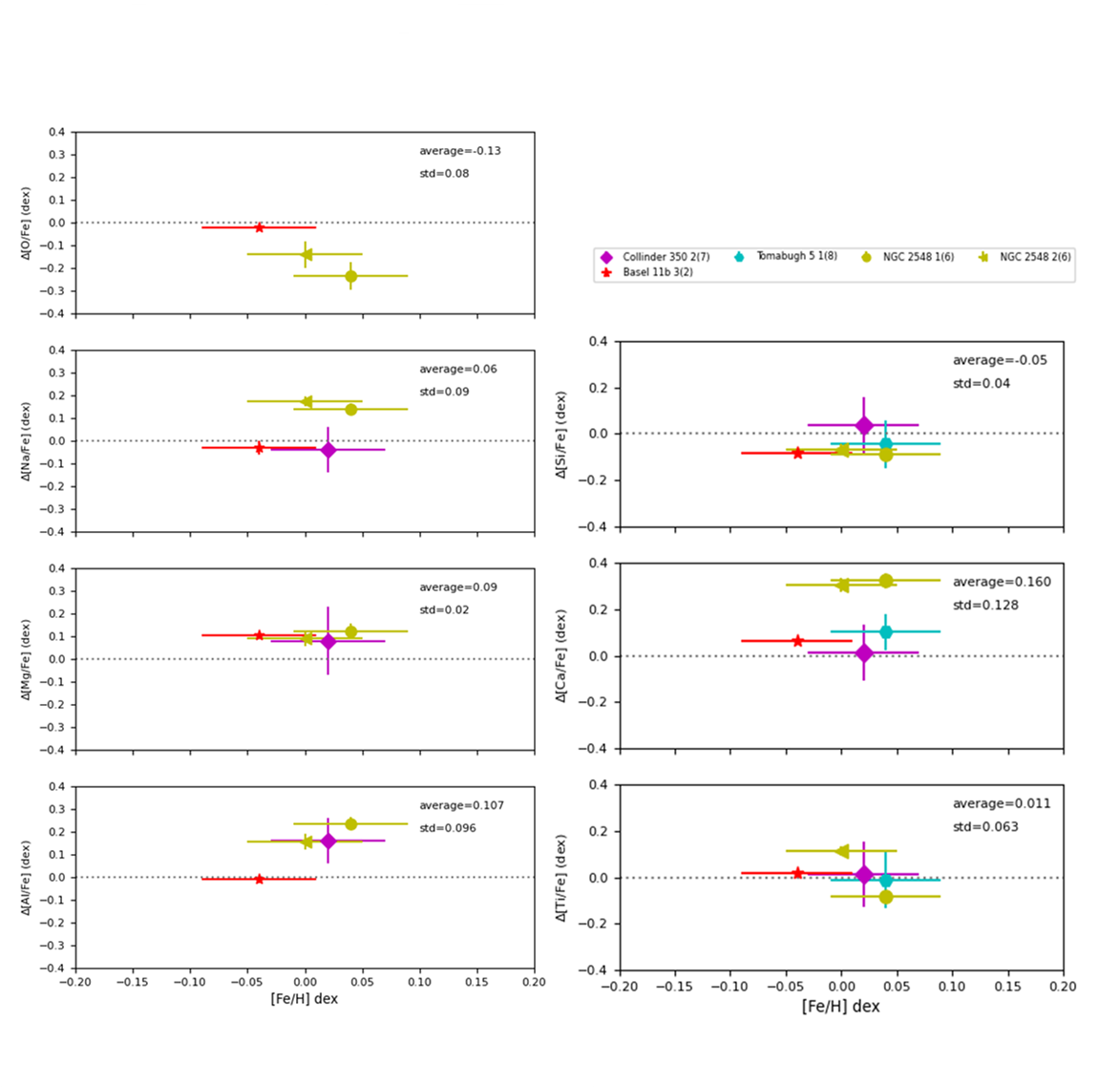}
     \caption{Comparison result of chemical abundance for young OCs with high-resolution determination. We plot our [Fe/H] in the x-axis, and the difference (our values minus the values in the literature) and the error from the literature on the y-axis. The numbers between parentheses close to the stars' names are the literature references, based on Table \ref{tab-comp}: (2)  APOGEE DR16; (6) \cite{spina21}; (7) \cite{casali20}; (8) \cite{baratella18}. 
     }
    \label{Figcomp1}
\end{figure*}

\begin{figure*}[htbp]
   \centering
    \includegraphics[height=18cm,width=18cm]{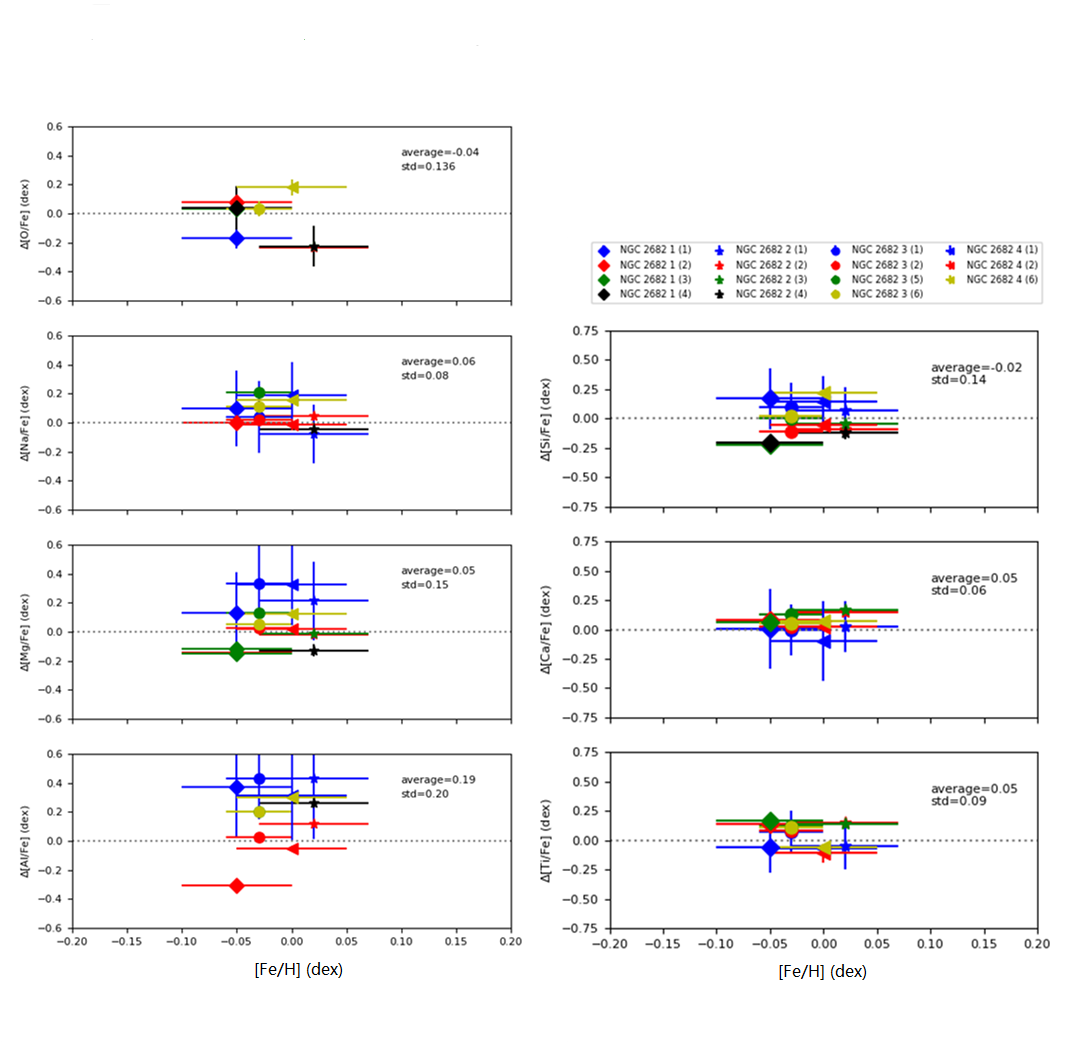}
    \caption{Comparison of chemical elements for targets in NGC 2682, The x-axis and y-axis are the same as in Fig.\ref{Figcomp1}. In the legend we show the different sources: (1) \cite{jacobson11}; (2) APOGEE DR16;  (3) \citep{casamiquela19}; (4) \cite{gao18}; (5) \cite{luck15}; and (6) \cite{spina21}}
         \label{Figcomp2}
   \end{figure*}

Figure \ref{Figcomp1} shows the comparison results for $\alpha$ elements, Na, and Al in young clusters, while Fig. \ref{Figcomp2} shows the comparison for the well studied NGC~2682, an older cluster. The relevant literature is listed in Table \ref{tab-comp}. 


For [O/Fe], we have a good agreement for Basel11b, but the comparison for NGC 2548 is less than optimal, hinting at a possible offset with \citep{spina21}. For NGC 2682, on the other hand, where more literature sources are present, there is a large scatter but no significant offset in Fig \ref{Figcomp2}. It is worth noticing, in this context, that \cite{spina21} used a combination of optical (GALAH) and IR (APOGEE) data, and both atomic and molecular features to derive the  abundances, which might have led to systematic effects in (some) abundances. Moreover, given the challenges discussed above that are associated with the derivation of O, some scatter is to be expected. 

In Fig \ref{Figcomp1} our Na abundances are slightly lower than those from \cite{spina21}, which, however, do not account for NLTE corrections. The average NLTE correction to the Na abundances for the two NGC~2548  stars in common with \cite{spina21} is $-$0.16\,dex, very similar to the Na offset abundance observed, which is $\sim-$0.18\,dex.   On the other hand, there is good agreement with Casali's result for Collinder350\_2, which did not take into account the NLTE correction. 
For NGC 2682, there are similar offsets in Na for studies that did not take into account NLTE corrections. There is, again, an offset between our result and those from  \cite{spina21}, and also similarly with \cite{luck15}  in Fig \ref{Figcomp2}, who also did not apply NLTE corrections. But the abundance of Na is compatible with the values reported by \citep{gao18}, who applied  NLTE corrections, but is interestingly also in good agreement with Jacobson's work, who does not.

For Mg, the comparison with the literature in Fig. \ref{Figcomp1} suggests the presence of an offset.
For Collinder350\_2 the difference might be due to differences in the atmospheric parameters  with respect to \cite{casali20}. The parameters differences are $\Delta T_{\textrm{eff}}$=130 K, $\Delta$log g=0.3 dex, and $\Delta$[Fe/H]=0.05 dex, which correspond to a change of $\sim$0.2\,dex, and the atomic parameters for the transitions are also slightly different, which log gf  on average 0.06 dex lower than ours. A similar offset is also observed  with respect to what was reported by APOGEE DR16 \citep{donor20} (or \citealt{spina21} who, indeed, combined GALAH and APOGEE clusters). 

However, in NGC 2682, our analysis is mostly in fair agreement with literature values, with the exception of \cite{jacobson11}, which have systematically higher abundances,  generally higher than what is reported in the literature.
We note that \cite{jacobson11} used moderate-resolution (R$\sim18000$) spectra and one single Mg line, at 6319 \AA, in a spectral range affected by telluric absorption features. The systematically high Mg abundance suggests the presence of blending that has not been accounted for. 
It is also worth noticing that \cite{jacobson11} used MARCS rather than Kurucz models. However, this should have led to abundance differences that are negligible with respect to the observational errors associated with the measurements under discussion.



For Al, we are in good agreement with APOGEE DR16
results (Basel11b and NGC 2682, \citealt{donor20}), but we find  various degrees of offset with respect to the works of \cite{spina21} (NGC 2548 and NGC 2682), \cite{jacobson11} (NGC 2682), \citet{gao18} (NGC 2682), and \cite{casali20} (Collinder 350). Some of them can be fully explained by the fact that they are LTE \citep{spina21,jacobson11,casali20}, and the NLTE correction can account for the differences. 
\citet{gao18}, on the other hand, did include NLTE corrections in their analysis, using the approach described in \cite{Nordlander2017}, the same that we adopt. The difference is therefore more likely to arise from the difference in atmospheric parameters (see Paper I) in the line list adopted and from the fact that Gao et al. used the GALAH pipeline, which performs an overall spectral fitting, rather than an individual lines analysis. 

Our values for [Si/Fe] are in good agreement with literature results for all targets in common. This indicates its suitability to be used (along with Ca and Ti, see below) as a probe of $\alpha$ elements in combined samples. 


The Ca abundance is in good agreement for all the stars in common with all literature sources, with the exception of \cite{spina21}, who report values  systematically larger for NGC 2548 and, to a lesser degree, for NGC 2682. As there are no systematics concerning the atmospheric parameters adopted (see \citealt{paperI}), we do not have any explanation for this, except a possible offset due to those abundances being the result of the combination of optical and IR data. 

Finally, the measured Ti abundances are in good agreement with the literature.
All these comparisons are shown in Figs.~\ref{Figcomp1} and \ref{Figcomp2}.

\section{Discussion}
As mentioned in the Introduction, the elemental abundances in the present sample provide information on different aspects. Lithium, Na, and Al are relevant in this context to shed light on stellar evolution and nucleosynthesis in young stars, providing constraints for models. 
On the other hand, $\alpha$-elements allow us to probe the characteristics of the Galactic thin disk.  


\subsection{Li content}

Lithium is destroyed at rather low temperatures, $\sim$2.5 $\times$ 10$^6$\,K, which roughly correspond to the temperatures at the base of the convective zone in  solar-mass stars. As stars evolve, Li gets depleted by mixing episodes (e.g., first dredge up) that mix the outer envelope with the interior.
Lithium abundance is usually lower than $\sim $1.5 dex for stars on the red giant branch (RGB) \citep[see e.g.,][]{casey16}. Lithium-rich giants have been detected both in the field and in clusters \cite[see e.g.,][where both open cluster and field stars where analysed homogeneously in the context of Gaia-ESO]{magrini21a}, but the reason for Li enrichment in these objects is still unclear. 

\begin{figure*}[htbp]
   \centering
   \includegraphics[height=7cm,width=20cm]{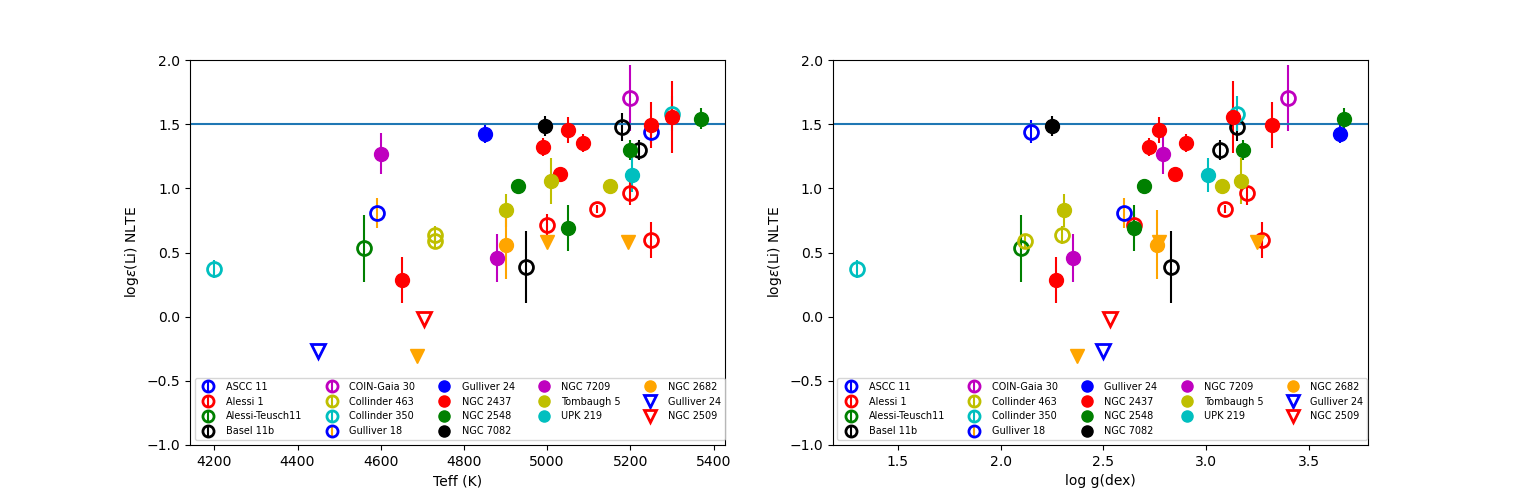}
      \caption{Distribution of Li abundance  versus T$_\textrm{eff}$ and log g for all samples. The inverted triangles show four stars with upper limit measurements, and the blue line is the value of a standard definition Li-rich giant, i.e., $\log \epsilon$(Li) = 1.5 dex.}
         \label{LiF}
   \end{figure*}

Figure~\ref{LiF} shows the behavior of NLTE $\log \epsilon$(Li) against T$_{\textrm{eff}}$ and $\log$g in our sample. Similar positive trends with LTE Li have been reported, for instance, by \citet{delgado16,magrini21b,franciosini20}. 
The four stars in our sample whose temperatures are higher than 5200~K (COIN-Gaia 30, Collinder 350\_2, NGC 2437\_3, and NGC 2548\_1) have $\log \epsilon$(Li) $>$ 1.5 dex, consistent with the result in Delgado's work from the hottest group of main-sequence stars (T$_\textrm{eff}$ $>$ 5000 K). 
We notice that some targets have a high Li abundance for their evolutionary stage. ASCC 11 and NGC7082\_2 have Li contents somewhat higher compared with the counterpart with a similar $\log$ g, and  Gulliver 37 and NGC 7209\_2 also have a slightly high lithium abundance for their effective temperature. 

\begin{figure}[htbp]
   \centering
   \includegraphics[height=6.5cm,width=9.5cm]{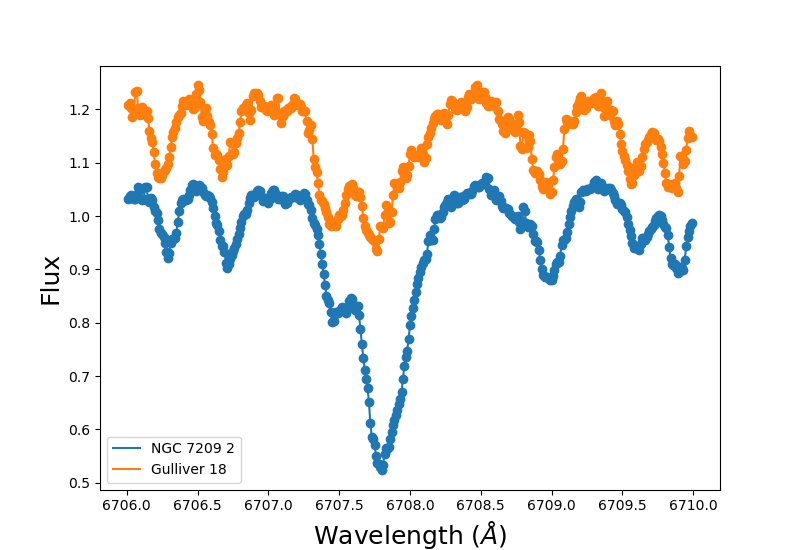}
      \caption{Comparison of  spectra around the 6707.81 $\AA$\, lithium line for two stars with similar stellar parameters and different Li abundances.
      They are the only stars observed in Gulliver\_18 ($T_{\textrm{eff}}$=4590~K, $\log$ g=2.60, [Fe/H]=-0.10~dex, log $\epsilon$(Li)=0.8) and NGC\_7209\_2 ($T_{\textrm{eff}}$=4600~K, $\log$ g=2.79, [Fe/H]=-0.07~dex, log $\epsilon$(Li)=1.27). }
         \label{overlap}
   \end{figure}

Our results show that our sample does not contain Li-rich stars. However, it is clear that scatter in Li abundance is present at all temperatures, as can be seen from Fig
\ref{overlap}, which exemplifies the observed differences, and shows the comparison of the spectral Li line for two stars with similar atmospheric parameters and different Li abundances.
This hints at other factors having an effect on the Li abundances in members of young clusters (e.g., age, metallicity, and binarity), consistently with what was discussed, for instance, by 
 \citet{gutierrez20}. 

\subsection{Sodium and aluminum content}
Sodium and aluminum are odd-Z elements,  mainly produced by Type II supernovae \citep{kobayashi20} and their production increases with increasing metallicity.
On longer timescales, low- to intermediate-mass stars, during their asymptotic giant branch (AGB) phase, can reach the temperatures to activate the NeNa and MgAl cycles, resulting in changes in Na and Al  on the stellar surface and ultimately contributing to the Galactic chemical evolution of these elements. 

\begin{figure*}[htbp]
   \centering
   \includegraphics[height=8cm,width=22cm]{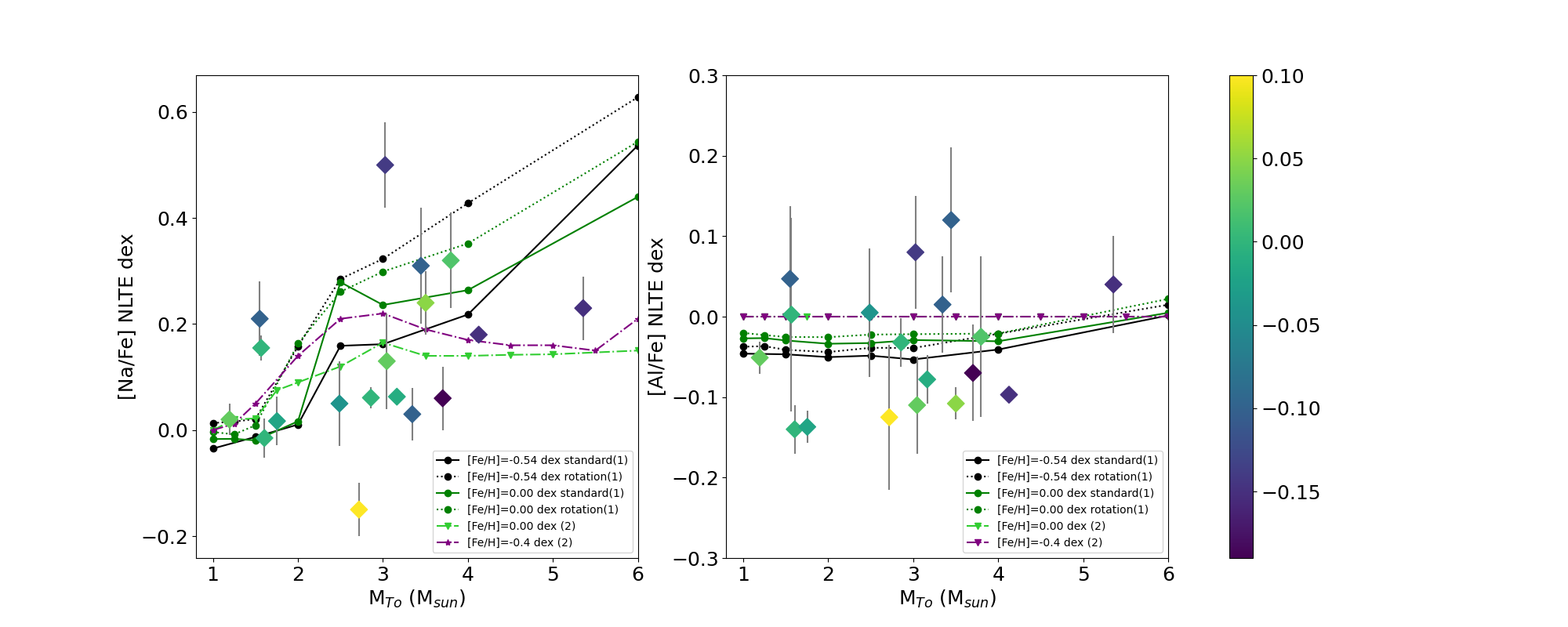}
      \caption{
       Comparison between the composition expected by stellar evolution models and  observation for the abundances of Na and Al (both in NLTE). The points are color-coded according to the cluster's metallicity, and the line colors also indicate the metallicity. References for models: (1) \cite{lagarde12}; (2) \cite{ventura13}. 
       }
         \label{mass}
   \end{figure*}

Therefore, Na and Al are of interest in order to probe both the stellar and Galactic chemical evolution. 
Several published works, such as \cite{jacobson07}, \cite{smiljanic12} and \cite{smiljanic16,smiljanic18} used the giants in OCs to explore the relationship between Na and Al abundances, and the first dredge-up.
Our sample is more numerous and has a somewhat wider age range, therefore allowing us to probe this issue on a wider stellar mass range.
Figure~\ref{mass} shows the comparison of our measurements for Na and Al with models.
Turn-off masses for each OC were derived using the isochrones of Padova (PARSEC release v1.2s, \citealt{2017ApJ...835...77M} and COLIBRI, \citealt{2020MNRAS.498.3283P}). The input age and metallicities of OCs are based on Table 1 and Table 4 in \cite{paperI}. 
The models are from \cite{ventura13}  ([Fe/H]$-$0.4 dex and 0) and from \cite{lagarde14}  ($-$0.54 and 0 dex) \footnote{The metallicities were derived assuming Z$_\sun$=0.014 from \cite{asplund09}.}. The \cite{lagarde14} models  are calculated with and without rotation, assuming, in the first case, a rotation that is initially 30\% of the critical velocity at the zero-age main sequence (see \cite{lagarde14}. The model predictions would be different for a different value of initial rotation, reflecting a different effect of the rotation-induced mixing. Therefore, at a given mass, the predictions are expected to show some scatter, the magnitude of which, however, cannot be estimated for the lack of available model predictions.
It should be noted that our comparison relies on the assumption that the stars had solar-scaled Na and Al during their main sequence phase. While we have no direct information about the composition of these stars during their main-sequence, Na and Al are known to be solar scaled among field disk dwarfs in the metal range relevant to this discussion (see e.g., the HARPS-N GTO sample, see Sect 3.3 and Fig. \ref{figX1}). 

The left panel of Fig.\ref{mass} shows the Na abundance as a function of turn-off mass. Models for low-mass stars, with M below $\sim$1--2 M$_\sun$, agree in predicting a modest, if any, Na variation in giants after the first dredge-up (as we observed in our OCs). Instead, [Na/Fe] changes are expected for more massive stars, with Na remaining at a constant (enhanced) level for M $>3$ M$_\sun$ in the \cite{ventura13} models, while continuing to increase with increasing mass in the \cite{lagarde14} models, with the uptick being more pronounced in the rotating models. 

Our results for Na span the same general range indicated by the models, but a detailed look at the plot provides some interesting information.
Indeed, the lowest-mass stars do not show any Na enhancement. However, already just below 2M$_\sun$, two of the clusters, NGC 2509 and Collinder 350, show a Na abundance higher than the predicted abundances. For both these clusters the measurements are based on a single star, but we note that in a previous study, \cite{smiljanic16} presented Na and Al abundances for six OCs (no overlap with our sample), finding a mild Na enhancement in a cluster with a turn-off mass of $\sim$1.2M$_\sun$.
For OCs with a turn-off mass larger than 2M$_\sun$, we observe enhanced Na content, even if with considerable scatter. 
In the 2-4 M$_\sun$ interval, the Na abundances span a larger range than that of the models. At the high end, the [Na/Fe]$=0.5$\,dex of ASCC 11  is  considerably higher than the \cite{ventura13} predictions, but could be likely accounted for by a Legarde model with a higher rotational velocity.
At the low end, there are several clusters with Na abundances that are below than any of the predictions. The most extreme case is Gulliver 37, where Na is actually under-abundant (the value is based on a single star, likely a binary, with [Fe/H]=0.1, and is the highest iron abundance in the sample), but several other clusters in this mass range have 0$<$[Na/Fe]$<$0.1, while even the lowest prediction, from the solar metallicity \cite{ventura13} model, are around 0.15-0.20 dex. 
This might suggest that the assumption of solar-scaled main-sequence Na abundance might not be appropriate. 

We note that \cite{smiljanic16}, who performed a similar study, did not find any clusters below the \cite{ventura13} predictions, nor did \cite{smiljanic18}, who added ten further clusters to the comparison. \cite{jacobson07} also measured Na and Al in three OCs, finding high values ([Na/Fe]$>0.4$\,dex) without NLTE corrections.
Above 4 M$_\sun$ we have one single cluster, NGC~7082, whose Na is in fair agreement with the \cite{ventura13} models, and not matched at all by any of the Legarde models.
The \cite{smiljanic18} sample has two clusters above 4 M$_\sun$, with two having a modest Na enhancement, similar to that observed by us in NGC7082, a reasonable match to the Ventura models, while the third seems to favor the Legarde tracks.

The right panel of Fig.~\ref{mass} shows the Al abundance versus turn-off  mass. 
The \cite{ventura13} models show no variation across the mass range considered, while the \cite{lagarde14} models predict a very modest increase above 4 M$_\sun$.
The observations show more scatter than the predictions, even if most of the measurements are characterized by rather large uncertainties. \cite{smiljanic16} present less scatter, but their abundances all lie above the Lagarde's and Ventura's model level. In particular, at odds with us, they find no subsolar Al in their clusters. We note, however, that they report LTE Al abundances, while we accounted for the NLTE effects, which are negative and can be quite large in stars such as those under discussion (see Section 2.4). However, \cite{smiljanic18} increased the sample and applied NLTE corrections, finding a somewhat larger scatter and a possible upturn at high masses (based on two clusters), but no evidence of subsolar Al in their clusters.



\subsection{Elemental ratios and the disk}
The $\alpha$-elements primary production sites are Type II supernovae, whose progenitors are massive stars ($\gtrsim$8-10 M$_\sun$). In old Milky Way populations, stars are generally enriched by  $\alpha$ elements, [$\alpha$/Fe]$=\sim 0.3-0.6$ dex, 
while in younger populations the situation is more nuanced, with generations of SNIa having increased the Fe content, decreasing the [$\alpha$/Fe] ratio \citep[see][and references therein]{matteucci21}.
Therefore, probing the $\alpha$-elements offers a way to gain insight into the history of the chemical enrichment of a population, and in this case, through the present sample, to investigate those processes in the thin disk.


Galactic chemical evolution models can be constrained by trends in the chemical abundances with Galactocentric radius and age, as well as by variations in these trends with time and with metallicity. 
Our SPA sample of OCs is distributed over the range 7700 $< R_{GC}<$ 10000 pc, and within 0.5 kpc from the disk, with ages from 40 Myr to 4.2 Gyr. The distribution, however, is not uniform, with 
most objects in the sample being young ($< 1.5$ Gyr ) and concentrated in the 8-10 \,kpc zone.

In the following subsections we discuss the distribution of $\alpha$-elements, as well as, Na and Al, with respect to their distances from the Galactic center and the Galactic plane, and age. Results are also complemented by measurements obtained by literature works based on other projects (APOGEE, Gaia-ESO, GALAH, OCCASO, see \citealt{paperI} for the appropriate references) and for field stars (APOGEE DR17,  \citealt{leung19} and the HARPS-GTO sample, presented in \citealt{adibekyan12}), and compared to the Minchev, Chiappini \& Martig (MCM) models (the chemo-dynamical thin-disk models by \cite{minchev13,minchev14}.) for the formation of the Galactic disk. 



\subsubsection{Abundance of clusters and field stars}
It is interesting to compare our results for OC stars to those derived for field disk stars.
The HARPS-GTO program provides high-quality spectra on dwarf field stars in the solar vicinity. We adopted the atmospheric parameters and chemical abundance derived by \cite{adibekyan12}, which provide measurements for all the elements relevant to the present paper, with the exception of O, and the age determined from \cite{delgado19}. We adopted the  \cite{minchev18} temperature, age and Mg quality cuts (5300-6000~K and $\delta$ [Mg/Fe] $<$ 0.07 dex, $\delta$ Age/Age$<$0.25 or $\delta$Age < 1 Gyr.),in order to select a reliable sample representing the local field. These were used for comparison purposes in \citet{casamiquela19}, and therefore the comparison is instructive, even if it must be kept in mind that systematic offsets might exist in the derived parameters and abundances.

\begin{figure*}
\centering
\begin{minipage}[t]{1\textwidth}
\centering
\includegraphics[width=20cm]{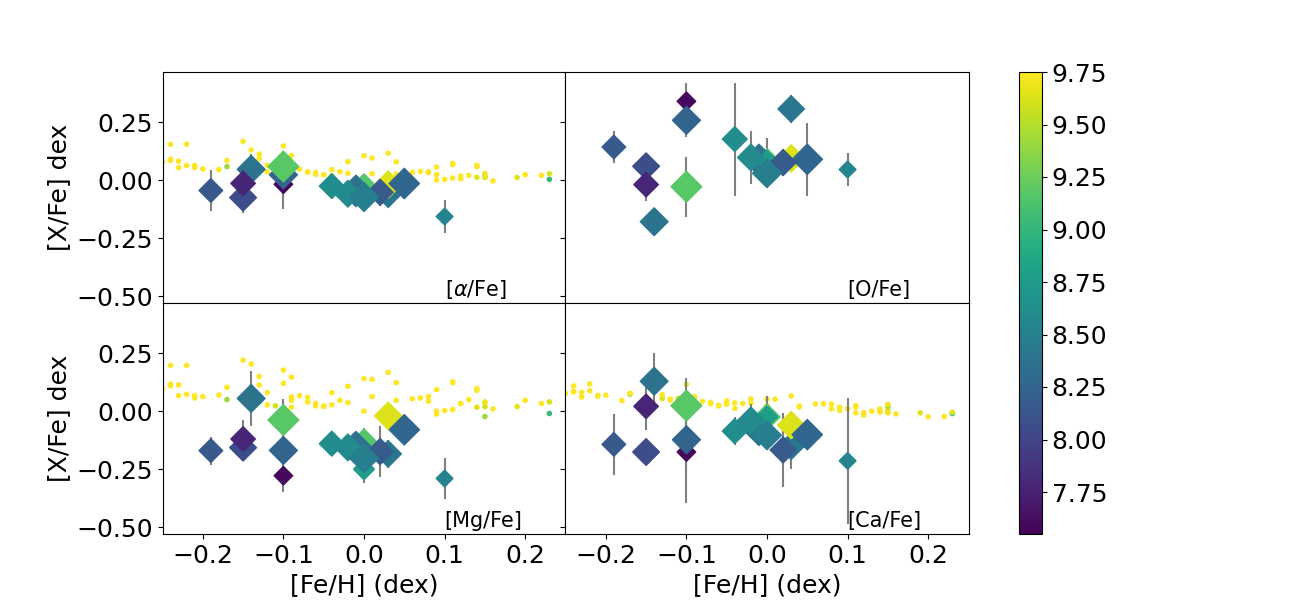}
\end{minipage}
\begin{minipage}[t]{1\textwidth}
\centering
\includegraphics[width=20cm]{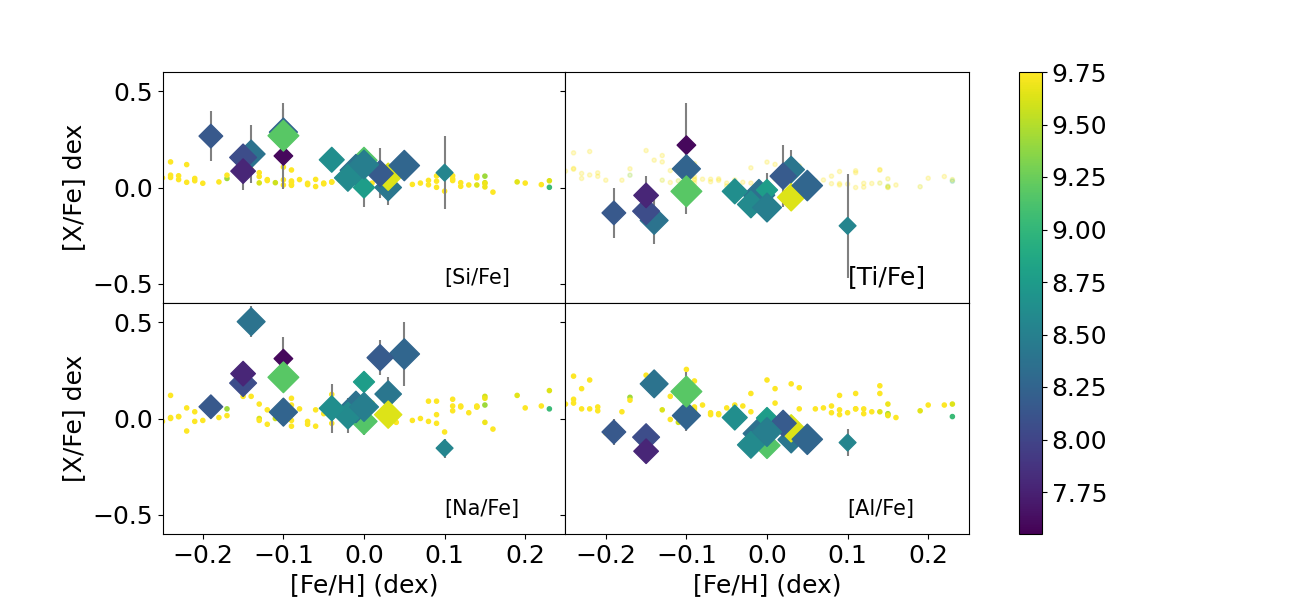}
\end{minipage}
   
     \caption{ Relationships between [Fe/H], the $\alpha$ elements (including [$\alpha$/Fe] defined as the average of Mg, Ca, Si, and Ti indexed against Fe), Na, and Al, colored by log (Age) and  sized by R$_{gc}$; the larger one is the sample with the longer R$_{gc}$.
     The small dots are for the HARPS-GTO samples. The O abundances are not available for the HARPS-GTO sample.
   The relationships between [Fe/H] and [Na/Fe] and [Al/Fe] are colored by log (Age). The small dots are for the HARPS-GTO samples.
}     
     \label{figX1}
   \end{figure*}

Figure \ref{figX1} shows the behavior of $\alpha$-elements, as a function of metallicity, in our clusters and in the HARPS-GTO field sample. Symbols are color-coded according to age. We plot O, Mg, Si, Ca, Ti, and the average of [$\alpha$/Fe], defined as the mean abundance of Si, Ca, Ti, and Mg weighted by the errors. 

Our clusters belong to the thin disk and are typically quite young,
while there is a general scarcity of stars of similar ages in the HARPS-GTO sample. The overall [$\alpha$/Fe] distribution is in reasonable agreement with the lower edge of the distribution of the field stars (which, as can be seen from the plot, corresponds to the younger objects, although these are still considerably older than the bulk of our sample). Magnesium and Ca have systematically lower abundance, while Si appears a little higher, and Ti in good agreement.

\cite{adibekyan12} adopted \cite{anders89} for the solar reference abundance, for which Mg, Si, and Ca are in excellent agreement with \cite{asplund09}.
A similar comparison was performed by \cite{casamiquela19} for the OCCASO data. They found a better agreement for Ca and Mg in the OCCASO sample, 
but their sample has an older age distribution, with more than 60\% of the clusters having Age $>$1 Gyr, while 15 out of our 18 clusters are below that age. It is worth noting, that the oldest clusters in our sample, Alessi 1, NGC 2509, and NGC 2682, with ages 1.5, 1.4 and 4.3\, Gyr, respectively, are generally in good agreement with the behavior of the oldest among the field stars.

The mismatch between the elemental ratios for Mg, Ca, Si, and Ti in the clusters and in the field could be due to the lack of stars of suitable age in the field sample, or to systematic effects related to the analysis of very young stars, or a combination of both. Paper I discusses the effects on Fe measurements, and this is also discussed in Sect. 3.4 regarding other elements.

The comparison with Na and Al shows a clear mismatch of Na, while for Al, cluster and field star are in fair agreement. We note, however, that Na is known to be affected by stellar evolution (see Section 3.2), with giants being enhanced to different degrees with respect to their main-sequence phase. Therefore, we expect that our sample will be characterized by a higher Na content with respect to dwarfs, which is consistent with the lower panel of Fig.~\ref{figX1}.

\begin{figure*}[htbp]
\centering
\begin{minipage}[t]{1\textwidth}
\centering
\includegraphics[width=22cm]{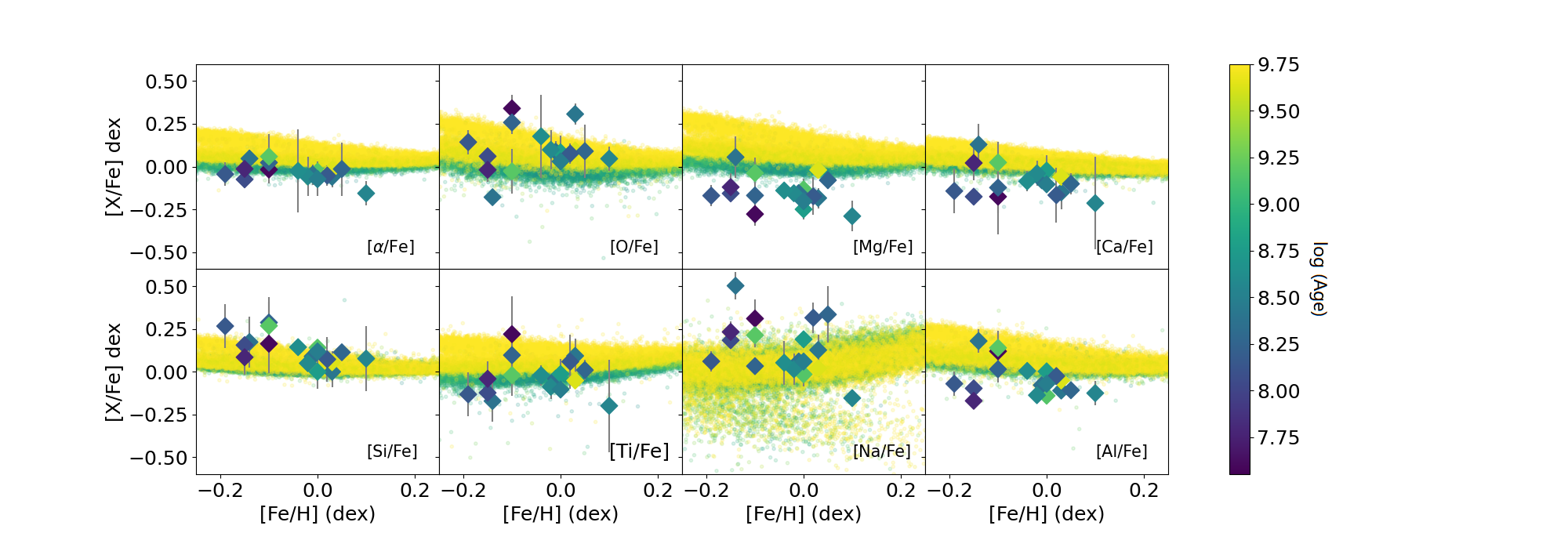}
\end{minipage}
\begin{minipage}[t]{1\textwidth}
\centering
\includegraphics[width=22cm]{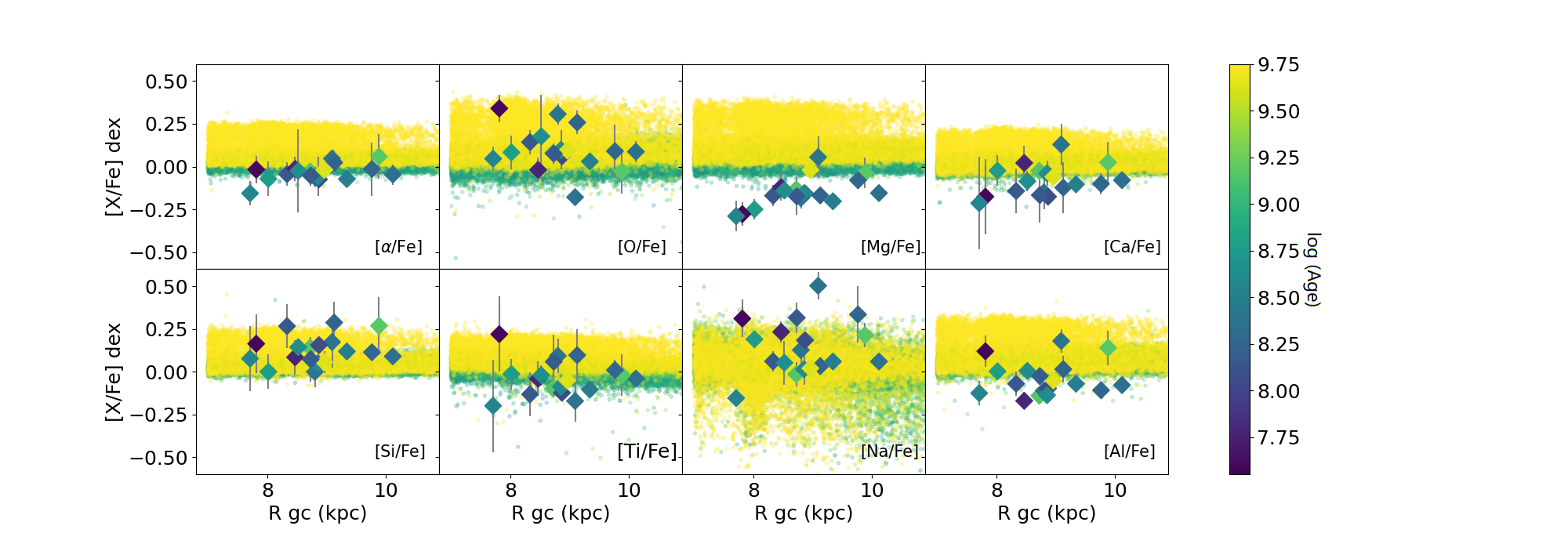}
\begin{minipage}[t]{1\textwidth}
\centering
\includegraphics[width=22cm]{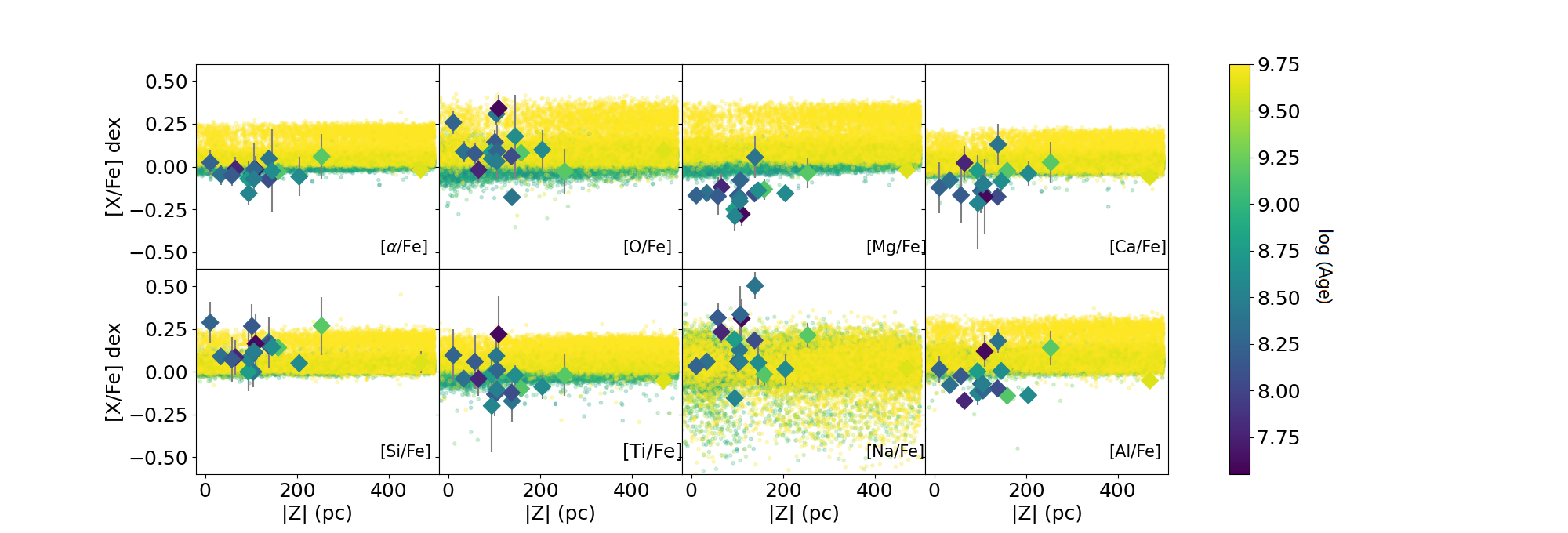}
\end{minipage}
\caption{Average [X/Fe] as function of [Fe/H], R$_{gc}$ and $|Z|$ for the SPA OCs,  colored by log (Age) and with uncertainties in chemical abundances indicated. The field giant stars (small dots, also colored with age) are selected from the APOGEE DR17. 
}
\label{fig_apogee}
\end{minipage}
\end{figure*}

The recent APOGEE data release contains an analysis for a large sample of young field stars, which might provide a more suitable comparison sample for our purposes. Figure \ref{fig_apogee} shows the measured elemental ratios with respect to metallicity, Galactocentric distance, and distance from the mid-plane. We used the latest result from APOGEE in DR 17 with distances and age estimates from the {\tt astroNN} catalog based on deep-learning code combined with multiple methods \citep{leung19}. We selected a sample of thin- disk giants (1.5 $<\log g <$ 3.5) spanning the same spatial range as our clusters,  |Z| < 0.5 kpc, 7 < R$_{gc}$ < 11 kpc. We also applied quality cuts for Mg and age analogous to those for the HARPS-GTO sample.
These criteria resulted in a sample of $\sim$67000 stars. 
\\
\\
The APOGEE DR17 thin-disk-giants sample is typically younger and shows a larger scatter in the abundances of all species compared to that observed for the HARPS GTO sample. 
Still, the thin-disk-giant sample contains just 300 stars below 500 \,Myr and just seven below 300\,Myr.

Overall, there is general agreement between the abundances measured in our sample and the APOGEE DR17 thin-disk giants, as can be seen from Fig.~\ref{fig_apogee}, especially when considering the youngest among the disk stars. 
We note, however, that Mg and and Ca are still lower in the clusters than in their field counterparts, an effect that also affects the average [$\alpha$/Fe].
The abundances of Mg and Ca seem to decrease in clusters closer to the Galactic center. This trend is expected by chemo-dynamical models (see sec 4.4), but interestingly it seems to not to be consistent with the abundances found in the field.
We note that \cite{casamiquela19} finds a discrepancy in the same direction in Mg their sample when comparing to APOGEE data. Similarly, \citet{magrini18}, reported an enhancement, but this is found for clusters in an inner Galactic position rather than the SPA OCs. 
For O and Si, some of the youngest clusters reach values higher than observed in the field, while Ti is in good overall agreement.The same is true for Na and Al, which ,however, have very large scatters in the APOGEE DR17, possibly due to evolutionary effects but also to more uncertain measurements. 


\subsubsection{Distributions of elemental ratios with respect to R$_{GC}$ and $|$Z$|$}


In order to explore the Galactic disk properties, probe its formation and provide constraints to model its formation, elemental ratios for OCs from multiple sources were collected in an extended sample, to cover a wider range of distances from the Galactic center and the disk, and of metallicity and ages, than available in our SPA sample. We included results from APOGEE DR16([X/Fe] and distances from \citealt{donor20}, and |Z| from \citet{cantat20}), Gaia-ESO ([X/Fe], distances, and |Z| from \citealt{casali19}),  GALAH ([X/Fe], distances and |Z|  from \citealt{spina21} for all clusters not already in the APOGEE sample), and OCCASO \citep{casamiquela19}. 
 

Furthermore, we included all the other clusters from the SPA project, as analyzed by \cite{dorazi20} \cite{frasca19}, \citet{casali20}, and \cite{alonso21}. The resulting collection is made up of a total of 152 clusters, covering the age range 4  Myr-7 Gyr, the Galactocentric range from  5.8 to 20 kpc, distance from the Galactic plane $|Z|$ up to 1750 pc, and a range from -0.5 to 0.4 dex in [Fe/H]. 

We note that in assembling the combined sample, we made no distinction between literature studies of giants and dwarfs. While this is not ideal in principle, the vast majority of the works considered ,in fact, targeted giants. Moreover, this approach is commonly adopted when deriving gradients.
It is important to keep in mind, however, that offsets related to analysis details are likely to exist between different sources regardless of whether the observed stars are giants or dwarfs, but at this stage this is the only way to assemble large samples of OCs, spanning age, distance and metallicity.
Upcoming large surveys that will uniformly target OCs (e.g., the WEAVE GA-OC survey and the 4MOST Stellar Cluster Survey) will yield a homogeneous analysis of several hundreds of OCs over the relevant parameter range, allowing gradients to be probed in much more detail.

\begin{figure*}
   \centering
   \includegraphics[height=22cm,width=15cm]{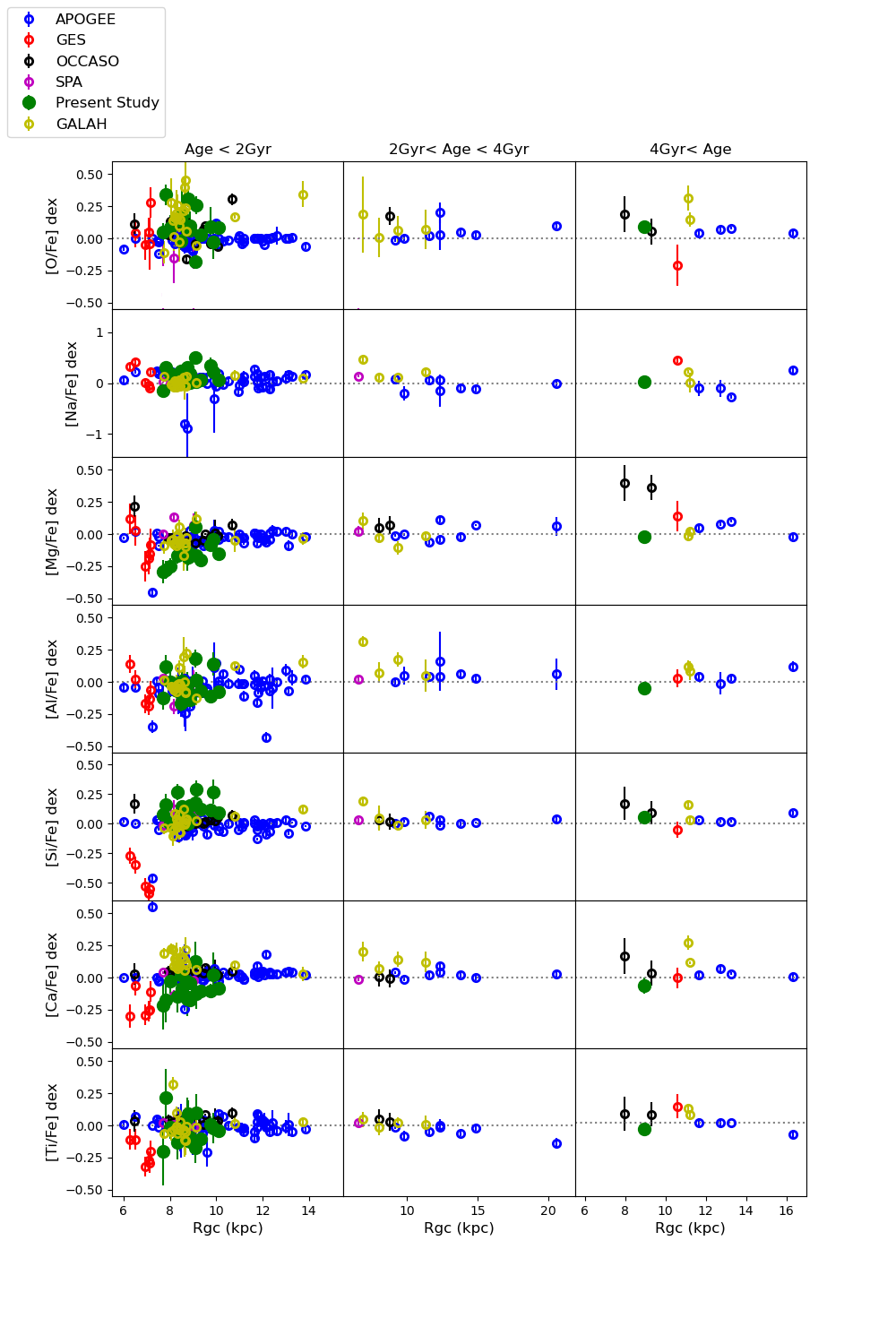}
  \caption{
      Distribution of abundance ratios with Galactocentric distance in three age bins. Besides our clusters, we show data from the APOGEE DR16\citep{donor20}, GES \citep{casali19}, OCCASO \citep{casamiquela19},and GALAH \citep{spina21}, plus SPA results already published  \citep{frasca19,casali20,dorazi20}.
      } 
         \label{fig_D}
   \end{figure*}
\begin{figure*}
\small
   \centering
   \includegraphics[height=22.5cm,width=15cm]{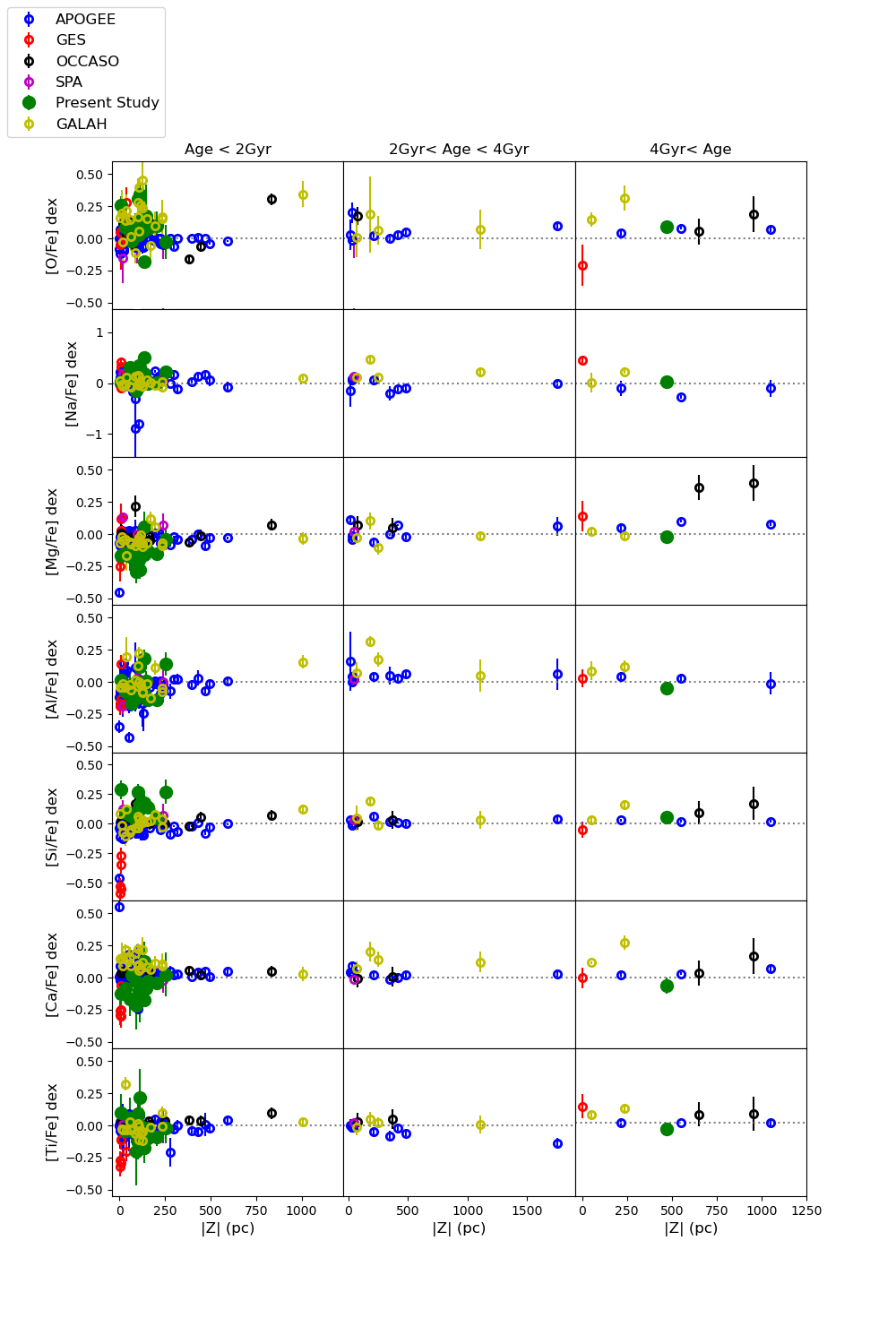}
     \caption{
      Distribution of abundance ratios with distance from mid-plate in three age bins. Besides our clusters, we show data from  the APOGEE DR16 \citep{donor20}, GES \citep{casali19}, OCCASO \citep{casamiquela19},and GALAH \citep{spina21} plus SPA results already published  \citep{frasca19,casali20,dorazi20}.
      }
         \label{fig_Z}
   \end{figure*}

Figures \ref{fig_D} and \ref{fig_Z} show the behavior of the measured abundance ratios with respect to R$_{GC}$ and $|$Z$|$, split into three age bins, which have been chosen to match the predictions of the  MCM models. 
Most of the objects in the combined sample are younger than 2 Gyr. 
The older age bins are populated mostly by APOGEE DR16 and GALAH OCs, with a small contribution from the present study (one cluster), Gaia-ESO (one cluster) and OCCASO (three clusters).
We note that in the youngest bin there is not only more scatter, but there are also systematic differences among different sources. This is discussed further in Section  \ref{sec_age}.



We derived the Galactocentric and vertical gradients of $\alpha$-elements, as well as Na, and Al
following the same approach in Paper I  concerning the grouping of clusters by distance and age\footnote{It should be note that for Si and Ca, Gaia-ESO clusters younger than 2 Gyr were excluded from the gradient derivation, as they seem to be affected by a negative offset (see next section for details). However, including them in the sample would not have affected the qualitative result.}.  Abundances for clusters that had measurements from more than one source were combined using a weighed mean before deriving the gradients.
Gradients were calculated using a linear fit, weighting according to the associated uncertainties. We  note that the combination of data from different sources is very likely to be affected by systematic errors in ages, abundance ratios, and distances. We further note that while we took the utmost care in assessing them and found them to be small with respect to the observational errors, there were only a limited number of stars and/or clusters in common that could be checked for offsets, covering only a fraction of the relevant parameter space.
Table \ref{tab:nagr} shows the gradients of [X/Fe] with distances. The bins for older clusters ($>$2 Gyr) and/or distant clusters are very scarcely populated, but we calculated the gradients in those bins for completeness.


Data for O, Na, Al, Si, and Ti are consistent with a flat distribution  with respect to Galactocentric distance and distance from the mid-plane. \cite{casamiquela19} reached the same conclusion for Si and Ti (they do not present the analysis for O, Na, and Al).
On the other hand, \cite{donor20},  using the APOGEE DR16 sample,  finds flat distributions for Si and Ti,  but small positive statistically significant gradients for O and Al, and negative for Na. It is worth noticing that the APOGEE DR16 sample has a different age distribution from the combined sample that we use, and it contains a larger fraction of very young clusters. However, we do not find any significant trend for O, Na or Al even when computing the trends on the basis of clusters older than 1 Gyr in the combined sample.

For Mg and Ca we derive statistically significant gradients, and in both cases their values are consistent when considering the whole sample or just clusters younger than 2\,Gyr, within the statistical errors.
The value for the radial gradient of Mg is in perfect agreement with that reported in \cite{casamiquela19}, 0.01$\pm$0.002 dex~kpc$^{-1}$, and \cite{donor20}, 0.009$\pm$0.001 dex~kpc$^{-1}$.
For Ca, \cite{casamiquela19} finds no significant gradient, while we find a significant positive trend, a finding similar to \cite{donor20}, who report an even steeper value (0.012$\pm$0.0001 dex~kpc$^{-1}$).
Even in this case, excluding the 18 clusters from this paper from the combined sample does not significantly affect the value.

\begin{figure*}
\centering
\begin{minipage}[t]{0.48\textwidth}
\centering
\includegraphics[width=9cm]{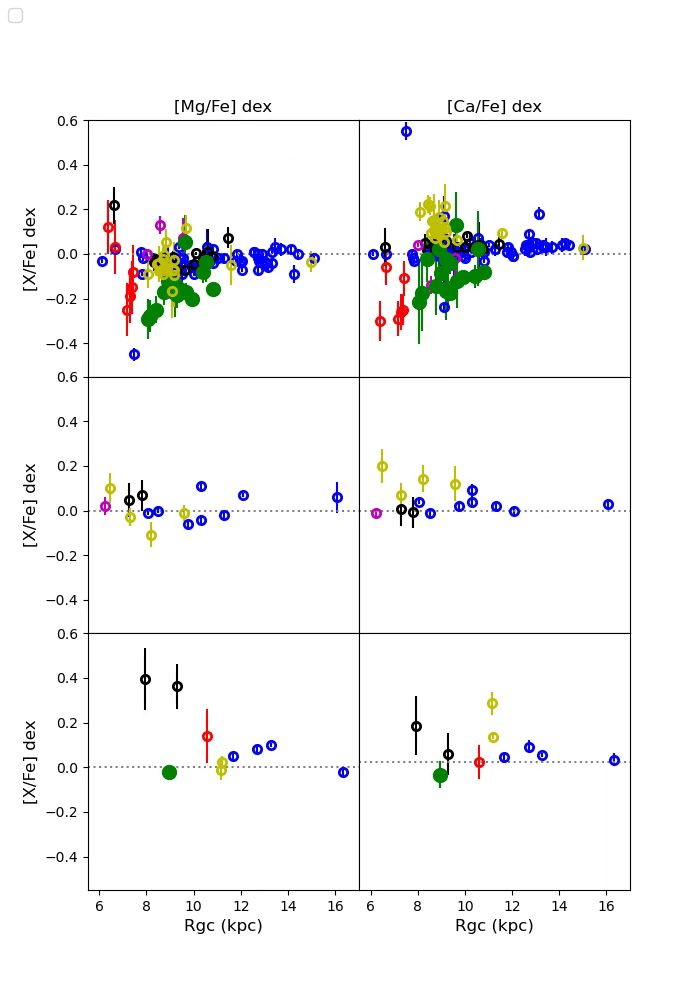}
\end{minipage}
\begin{minipage}[t]{0.48\textwidth}
\centering
\includegraphics[width=9cm]{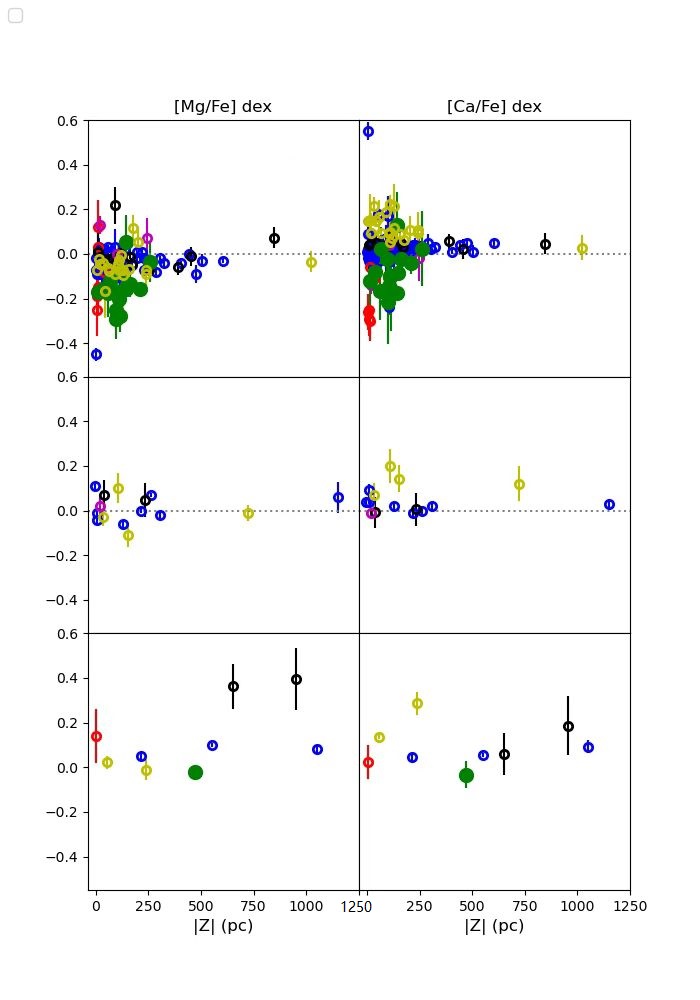}
\end{minipage}
   
     \caption{ Enlarge plot of Mg and Ca based on Figs.
    \ref{fig_D} and \ref{fig_Z}
}     
     \label{figD1}
   \end{figure*}

For the gradients with distance from the mid-plane, we also find significant trends only in Mg and Ca. The enlarged figure is Fig \ref{figD1}. 
\cite{boeche14} derived vertical gradients for $\alpha$ elements on the basis of a sample of field red giants from RAVE, and \cite{hayden15} used APOGEE data to investigate $\alpha$ content at different heights on the Galactic plane. They both find that the $\alpha$ increases with increasing |Z|, consistently with the gradients we determine for Mg and Ca.

We note that this is likely related to a variation in the age distribution as a function of distance from the Galactic midplane. In fact, at any distance from the Galactic plane, the predictions for a given age are the same and the vertical gradients observed are likely due to a larger fraction of older stars encountered with increasing |Z|.

\begin{table*}[ht]
\setlength{\tabcolsep}{1.25mm}
\begin{center}
\caption{Observed gradients of $\alpha$-elements, and Na and Al (with NLTE correction).} 
\begin{tabular}{lrrrr}
\hline\hline
Age range& R$_{gc}$ & d[O/Fe]/dR$_{GC}$ & d[O/Fe]/d$|z|_{GC}$&N$_{clusters}$\\
(Gyr) & (kpc) &  (dex~kpc$^{-1}$) &(dex~kpc$^{-1}$)& \\
\hline
all ages& Rgc<14&$0.0030\pm0.0060$  &$0.0506\pm0.0271$ &111  \\
 age<2& Rcg<14&$0.0506\pm0.0271$    &$-0.0261\pm0.0266$  &91  \\
 2<age<4 & Rgc<14&$0.0121\pm0.0028$ &$0.0943\pm0.0467$  &13   \\
 age>4& Rgc<14& $0.0006\pm0.0160$ & $0.0130\pm0.0531$ &9   \\
 \hline\hline
Age range & R$_{gc}$ & d[Na/Fe]/dR$_{GC}$ & d[Na/Fe]/d$|z|_{GC}$&N$_{clusters}$\\
(Gyr) & (kpc) &  (dex~kpc$^{-1}$) &(dex~kpc$^{-1}$)& \\
\hline
all ages& Rgc<14&$0.0013\pm0.0052$ &$0.0026\pm0.0057$  &117  \\
 age<2& Rcg<14&$0.0064\pm0.0058$ &$0.0290\pm0.0676$ &99 \\
2<age<4 &Rgc<14&$0.0227\pm$0.0103 & 0.0271$\pm$0.1176  &11  \\
age>4& Rgc<14& $-0.1041\pm$0.0635 & $-$1.0445$\pm$0.2437& 5   \\
 \hline\hline
 Age range & R$_{gc}$ & d[Mg/Fe]/dR$_{GC}$ & d[Mg/Fe]/d$|z|_{GC}$&N$_{clusters}$\\
(Gyr) & (kpc) &  (dex~kpc$^{-1}$) &(dex~kpc$^{-1}$)& \\
\hline
all ages& Rgc<14& $0.0099\pm0.0017$ &$0.0703\pm0.0191$ &144  \\
 age<2& Rcg<14& $0.0095\pm0.0017$  &$0.0340\pm0.0244$  &122  \\
 2<age<4 & Rgc<14&$-0.0052\pm0.0050$ &$-0.0024\pm0.0410$  &13   \\
 age>4& Rgc<14&$0.0270\pm0.0093$&$0.0713\pm0.0708$ &9   \\
 \hline\hline
 Age range& R$_{gc}$ & d[Al/Fe]/dR$_{GC}$ & d[Al/Fe]/d$|z|_{GC}$&N$_{clusters}$\\
(Gyr) & (kpc) &  (dex~kpc$^{-1}$) & (dex~kpc$^{-1}$)&\\
\hline
all ages& Rgc<14&$0.0074\pm0.0039$ &$0.0961\pm0.0354$ &127  \\
 age<2& Rcg<14&$0.0079\pm0.0031$   &$0.1017\pm0.0406$  & 109 \\
 2<age<4 & Rgc<14&$0.0037\pm0.0029$ & $ 0.0941\pm0.0696$  &11   \\
age>4& Rgc<14& $ 0.0188\pm0.0075$ &$-0.0741\pm0.0737$ & 7  \\
 \hline\hline
 Age range& R$_{gc}$ & d[Si/Fe]/dR$_{GC}$ & d[Si/Fe]/d$|z|_{GC}$&N$_{clusters}$\\
(Gyr) & (kpc) &  (dex~kpc$^{-1}$) & (dex~kpc$^{-1}$)&\\
\hline
all ages& Rgc<14& $-0.0011\pm0.0017$&$0.0148\pm0.0157$  &135  \\
 age<2& Rcg<14&$-0.0027\pm0.0019$   &$-0.0092\pm0.0266$  &113  \\
2<age<4 & Rgc<14&$-0.0035\pm0.0026$ &$0.0214\pm0.0393$  &13   \\
 age>4& Rgc<14&$-0.0089\pm0.0030$&$-0.0142\pm0.0211$ &9   \\
 \hline\hline
 Age range& R$_{gc}$ & d[Ca/Fe]/dR$_{GC}$ & d[Ca/Fe]/d$|z|_{GC}$&N$_{clusters}$\\
(Gyr) & (kpc) &  (dex~kpc$^{-1}$) &(dex~kpc$^{-1}$)& \\
\hline
all ages& Rgc<14& $0.0055\pm0.0015$  & $0.0424\pm0.0188$ &135  \\
 age<2& Rcg<14&$0.0050\pm0.0018$& $0.0547\pm0.0250$ &113  \\
 2<age<4 & Rgc<14& $0.0026\pm0.0049$& $-0.0392\pm0.0805$ &13   \\
 age>4& Rgc<14&$-0.0282\pm0.0224$&$ -0.1015\pm0.0664$ &9   \\
 \hline\hline
 Age range& R$_{gc}$ & d[Ti/Fe]/dR$_{GC}$ & d[Ti/Fe]/d$|z|_{GC}$&N$_{clusters}$\\
(Gyr) & (kpc) &  (dex~kpc$^{-1}$) & (dex~kpc$^{-1}$)&\\
\hline
all ages& Rgc<14& $-0.0033\pm0.0021$  & $0.0055\pm0.0222$ &150  \\
age<2 & Rcg<14 & $-0.0027\pm0.0025$ & $0.0277\pm0.0299$  & 116  \\
 2<age<4 & Rgc<14& $-0.0104\pm0.0029$& $-0.0767\pm0.0413$ &13   \\
 age>4& Rgc<14&$0.0030\pm0.0198$&$-0.0639\pm0.0587$ &8   \\

\hline
\end{tabular}
\label{tab:nagr}
\end{center}
\end{table*}

\subsubsection{Elemental ratios and age \label{sec_age}}
Ages can be reliably determined in OCs, unlike in field stars. Therefore, they are ideal tools for investigating  the chemical evolution of the disk. In this section, we explore the chemical trend with ages through the combination samples.
To ensure homogeneity, all the ages for the SPA OCs  and for the combined sample are from a single source: \cite{cantat20}.

\begin{figure*}
\small
   \centering
   \includegraphics[height=22.5cm,width=15cm]{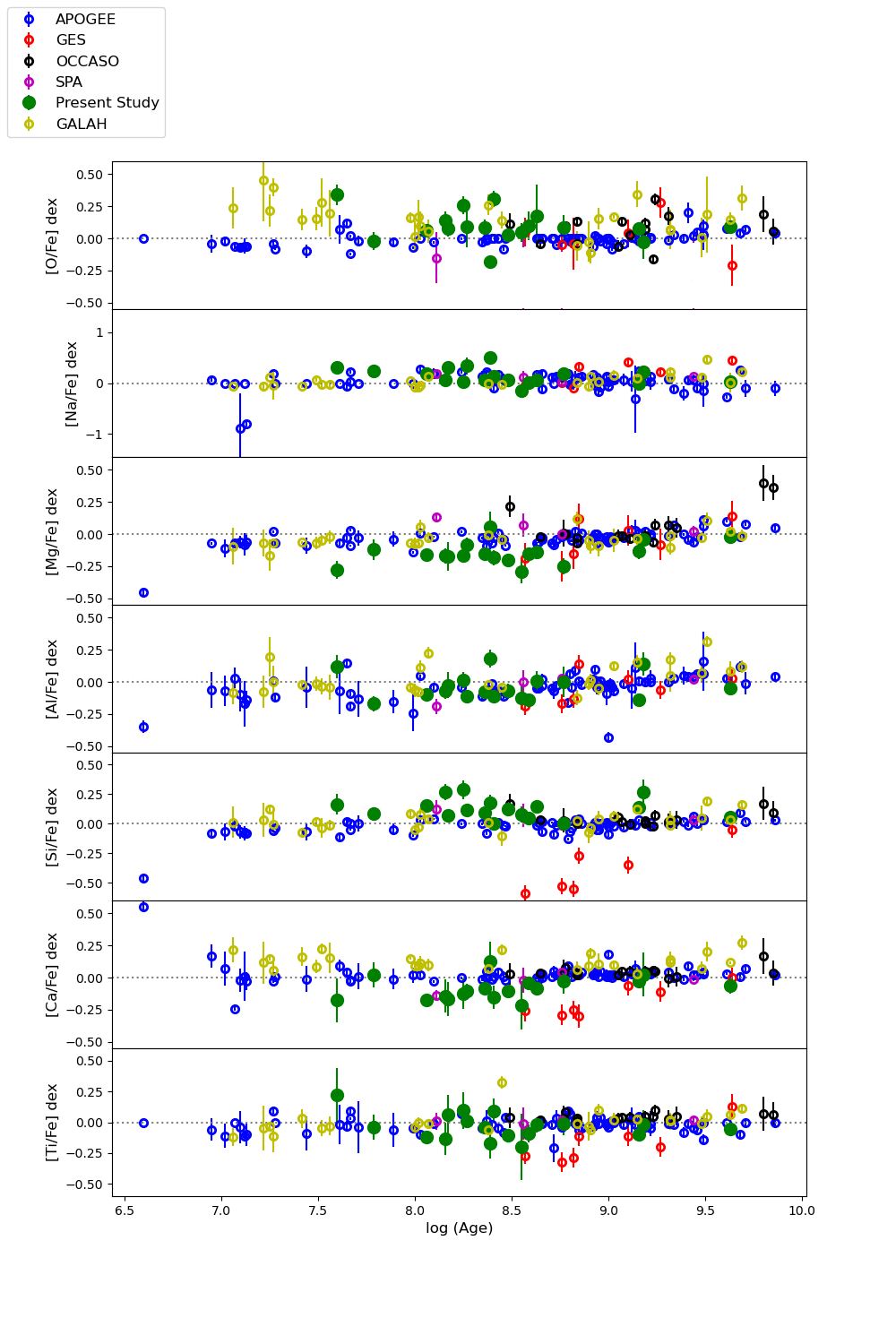}
     \caption{
         Distribution of abundance ratios with age. Besides our clusters, we show data from the APOGEE DR16\citep{donor20}, GES \citep{casali19}, OCCASO \citep{casamiquela19},and GALAH \citep{spina21}, plus SPA results already published  \citep{frasca19,casali20,dorazi20}.}
         \label{fig_age}
   \end{figure*}

Figure~\ref{fig_age} plots the abundance ratios of chemical abundance as a function of OCs' ages, colored according to their source.
We note that for some elements, the behaviors are quite different depending on their source.
 Aluminum and Na appear somewhat lower in APOGEE DR16 young clusters, an effect that might be due to the fact that both are derived without accounting for NLTE effects.
 For Si and Ca, Gaia-ESO reports abundances in very young clusters that are ,on average, lower than those in the other works considered here.
 The average Gaia-ESO abundances for Si and Ca for clusters younger than 2 Gyr are $-0.46\pm0.14$ and $-0.23\pm0.09$, respectively. This has to be compared with the abundance from OCCASO (0.03$\pm$0.05 and 0.04$\pm$0.01, respectively), the APOGEE DR16 ($-0.02\pm0.07$ and 0.02$\pm$0.08, respectively), GALAH ($-0.02\pm0.04$ and 0.09$\pm$0.05, respectively) and the SPA sample (0.13$\pm$0.08 and $-0.08\pm0.09$, respectively). Indeed, there is a considerable systematic difference, which is not observed for other elements. 
The disagreement seems much less severe among older clusters (age > 2 Gyr) and in particular all sources are in very good agreement for keystone clusters, such as NGC 2682 (keeping in mind the different approaches to the treatment of NLTE effects for Na and Al).

Even taking into account these offsets, and hence not considering the Gaia-ESO data for Si and Ca, it is clear that the scatter in the elemental ratio increases with decreasing age for all elements. This is expected to some extent: very young clusters are expected to have different compositions depending on their birth position, with clusters at large R$_{GC}$ being typically more metal poor and having a higher [$\alpha$/Fe]  with respect to  those forming in the inner disk, see e.g., \citet{chiappini09} and Figure 12 in \citet{casamiquela19}. However, the effect should be of rather modest magnitude, smaller than what can be noted from Fig.~\ref{fig_age}, suggesting the existence of some other cause.

On the other hand, the disagreement could  be related to the fact that traditional 1D analysis is inadequate for very young stars, as discussed in Paper I for the case of Fe, (see also the next section for further discussion).
However, we note that the differences extend beyond the range expected to be affected by these issues, hinting at the fact that the derivation of abundances in young stars might be particularly sensitive to the detailed assumptions made by different approaches to the analysis, including the choice of model atmospheres, the path to the derivation of the atmospheric parameters, and the specific transitions on which the abundance measurements rely.


We derive the gradients with age for the $\alpha$-elements, and for Na and Al, for the combined sample, and for just our 18 clusters. Gradients are listed in Table \ref{tab-age}. 
In this context, it is important to keep in mind that the SPA sample is heavily biased toward young clusters, lessening the significance of age gradients based just on such samples. 


Data are consistent with a flat distribution with age for Na, Si, and Ti, both for the combined sample and for just the 18 clusters presented in this paper.
For O and Al, the combined sample shows a statistically significant positive gradients. For Mg and Ca, positive gradient are present both in the combined sample and in the sample of the 18 present clusters. 

Our findings are in qualitative agreement with \cite{casamiquela19}, who do not report the values of the derived trends but show the associated plots. 
\cite{yong12} measured the age gradients for a number of $\alpha$, Fe-peak elements, and n-capture elements, finding a significant gradient for Mg, even if flatter than ours, but not for Ca. They also found significant trends for Na and Al, derived without accounting for NLTE effects.


\begin{table}[ht]
\setlength{\tabcolsep}{1.25mm}
\begin{center}
\caption{Chemical gradients}
\begin{tabular}{lrrr}
\hline\hline
    & Combined sample & &Present sample\\             
Species   &d[X/Fe]/d(Age)& n &d[X/Fe]/d(Age)\\
  & (dex/Gyr)&& (dex/Gyr)\\
\hline
O    & 0.018$\pm$0.004&114&0.008$\pm$0.011  \\
Na    &0.013$\pm$0.013&120&-0.031$\pm$0.040 \\ 
Mg & 0.029$\pm$0.004 &147&0.036$\pm$0.007 \\ 
Al	 & 0.027$\pm$0.005&130& 0.009$\pm$0.008  \\ 
Si & 0.007$\pm$0.002&135&-0.014$\pm$0.006  \\ 
Ca & 0.013$\pm$0.003&135&0.082$\pm$0.028 \\
Ti & 0.000$\pm$0.004&140&0.002$\pm$0.010\\
\hline
\end{tabular}
\label{tab-age}
\end{center}
\tablefoot{The chemical gradient for the combined sample (second column) along with the number of clusters it is calculated on. The gradient for the 18 SPA OCs in the present paper is given in the last column.}
\end{table}

 \subsection{Comparison with chemo-dynamical models}
 
 %


Comparison of the measured abundance ratios with theoretical predictions provides insight into the processes that led to the formation of the disk.
In this context, we used the state-of-the-art MCM chemo-dynamical models. They combine chemical evolution from \citet{chiappini09}, describing how the composition of stellar populations change with time, with the movement and mixing processes affecting stars and gas, including stellar migration and mergers. 

Surveys of the disk field stars and large samples of stellar clusters, such as those in our combined sample, generally span considerable ranges in distance, height on the Galactic mid plane and ages, and measured trends and gradients are affected by the motions experienced by stars and clusters. Therefore, the combination of dynamics and chemical evolution in the models is crucial for a meaningful comparison with the observations.

\begin{figure*}[htbp]
   \centering
    \includegraphics[height=18cm,width=18cm]{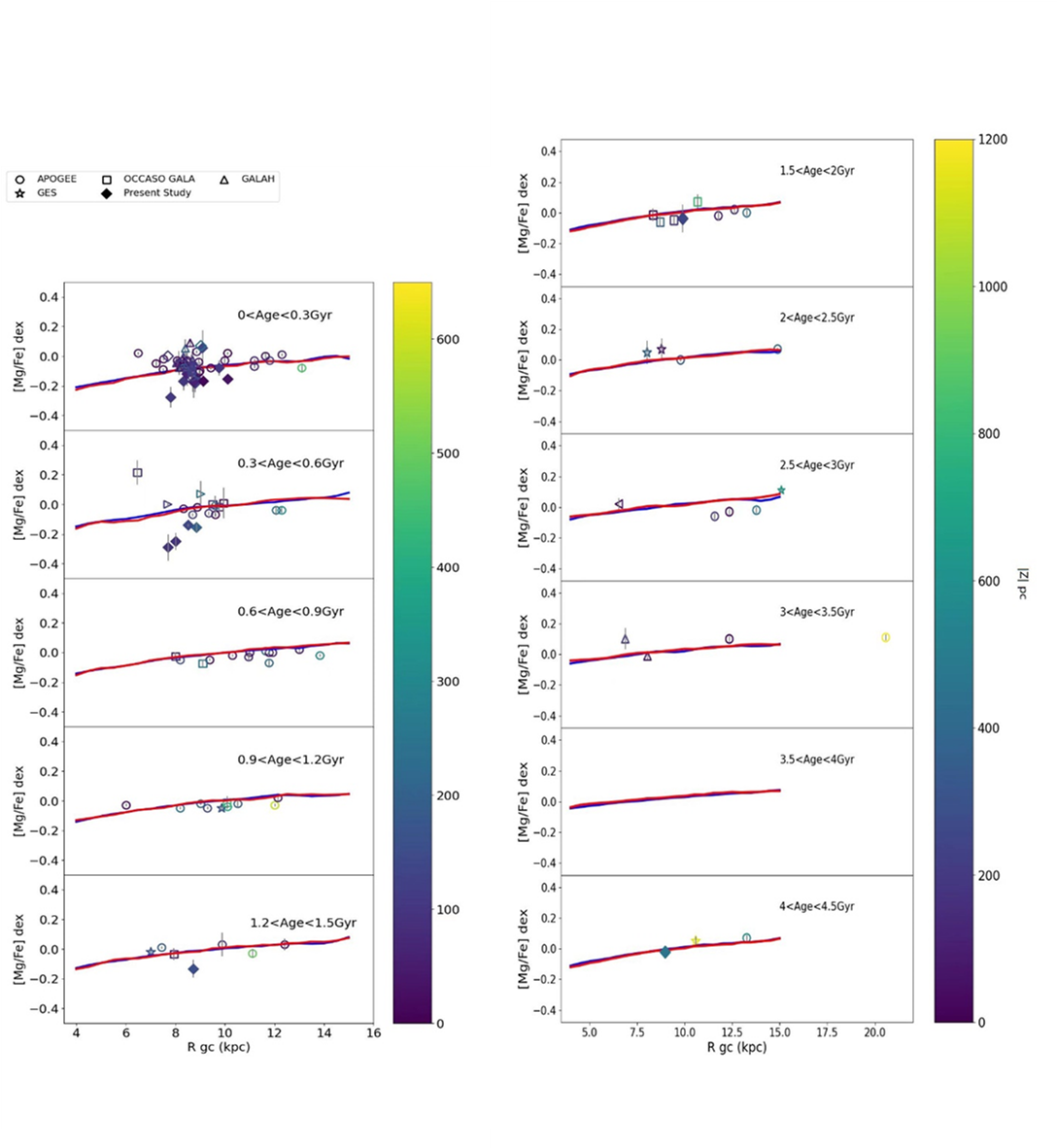}
    \caption{Comparison between model predictions of magnesium \citep{minchev13,minchev14} and observations for clusters younger than 1.5 Gyr. The red and the blue lines are simulations from the MCM models for $|z| < 0.3$\,kpc and 0.3 $< |z| < $0.8 kpc respectively. The color in the symbol indicates the distance from the Galactic midplane. All clusters are within 0.6 kpc from the Galactic plane, and SPA clusters are all within 0.5\,kpc. The open "diamond" in the first panel is ASCC 123 \citep{frasca19} and the "triangle" symbol in the 0.3-0.6 Gyr range is NGC 2632 \citep{dorazi20}, for clusters between 1.5 and 4.5 Gyr. Only one SPA cluster is older than 4 Gyr:  Ruprecth~171, from \citealt{casali20})}
         \label{fig_mg1}
   \end{figure*}

\begin{figure*}[htbp]
   \centering
    \includegraphics[height=18cm,width=18cm]{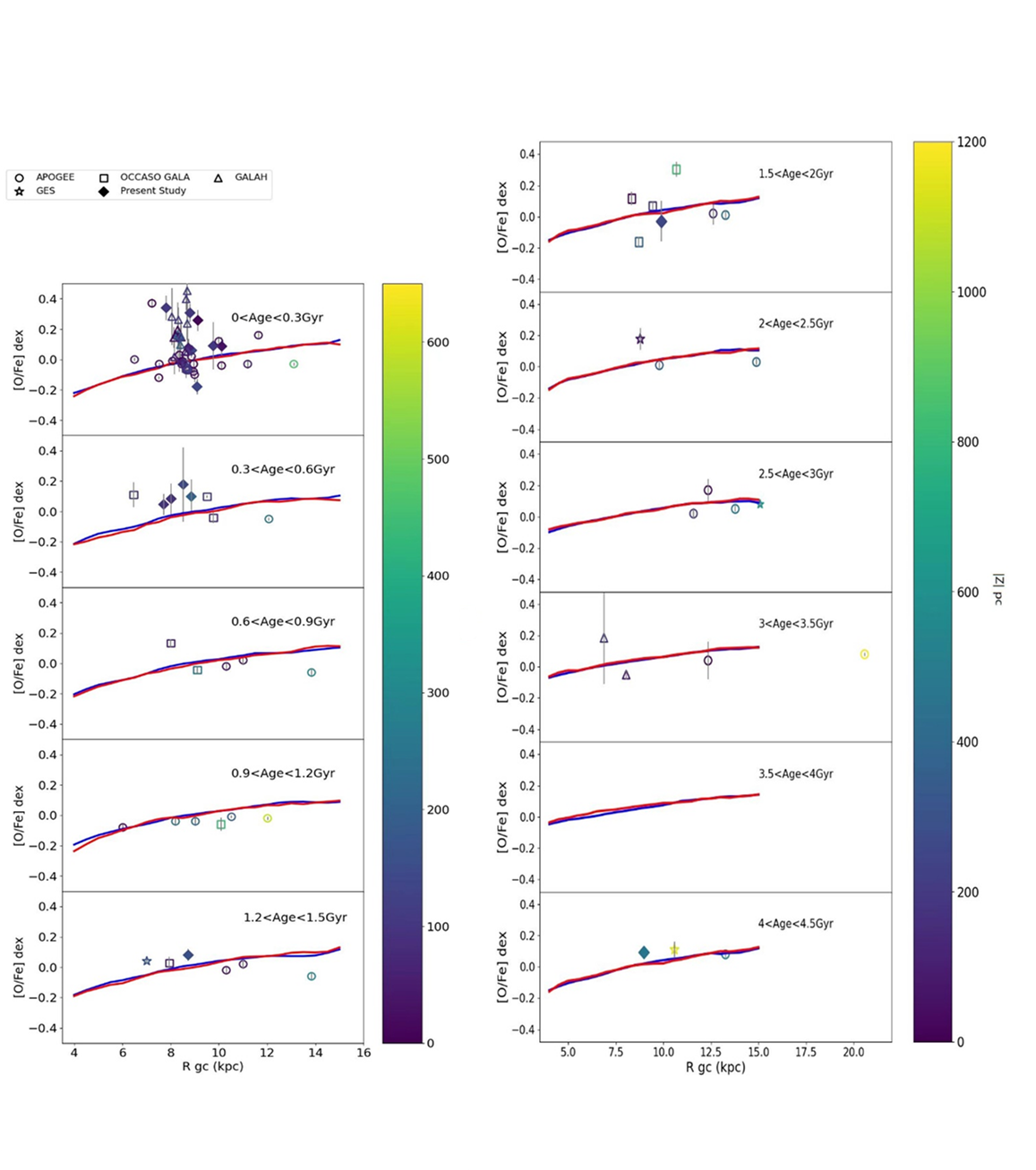}
    \caption{Comparison between model predictions of oxygen \citep{minchev13,minchev14} and the observations for young samples. The colors and symbols are the same as for as Fig.~\ref{fig_mg1}.}
         \label{fig_o1}
   \end{figure*}

In the following plots, the 
MCM models of the thin disk 
are rescaled so that the most likely birth position for the Sun (R$_{GC}$ = 6 kpc, 4.5 Gyr ago) matches the solar composition (see Minchev et al. 2013).
Figures \ref{fig_mg1} and \ref{fig_o1} show the comparison  of observational data with the MCM model predictions for [Mg/Fe] and [O/Fe] in the age range from 0-4.5 Gyr. The age bins are 0.3 Gyr for the younger ages (up to 1.5 Gyr) and 0.5 Gyr for older objects. The MCM models are  computed at two different distances from the midplane: |Z| < 0.3 kpc (blue lines) and 0.3< |Z| < 0.8 kpc (red lines). 

We limit our comparison to O and Mg, as those are the only ones fully published at this time.
Magnesium seems to be well reproduced overall by models. For young clusters (age < 0.6 Gyr), 
the data show a higher Mg content than predicted by the MCM models. It is worth noticing that our data (the 18 SPA OCs) are actually better reproduced by the models than the literature values (see upper left panel in Fig.~\ref{fig_mg1}), which report Mg abundances higher than the predictions for clusters younger than 0.6\,Gyr.
The vertical Mg gradient for the MCM simulation for clusters with age $<$ 2\,Gyr, 0.028 dex~kpc$^{-1}$, is in good agreement with the value calculated on the basis of the young clusters in the combined sample, 0.034$\pm$0.024 dex~kpc$^{-1}$. 

For O, the match of predictions is poorer, there is considerable scatter in the youngest bin, and generally the O abundances for our SPA sample and the literature are higher than what the models predict.
As mentioned in Section \ref{sec_age}, the discrepancy between observations and model predictions, and the offsets among different sources, are found in particular among very young clusters.

In Paper I we discussed the issues in measuring Fe abundances in very young stars, likely related to effects that are generally neglected in the modeling of stellar atmospheres (e.g., chromospheric activity and magnetic fields). These effects are discussed in the literature as being at play in very young dwarfs \citep{yana19, baratella2020, spina20}, making the derivation of atmospheric parameters through the traditional 1D LTE analysis based on minimizing trends for Fe lines inadequate. 
Our sample showed how these effects extend to giants (see Paper I) concerning iron. It is thus reasonable to expect that similar issues would plague the derivation of other species, providing a possible explanation for the large scatter observed among the youngest objects (age below $\sim$200\,Myr). Our data confirm this expectation.

This means that the mismatch between models and observations in very young clusters is of scarce astrophysical meaning; significant insight into the very young disk will require the application of a different path to abundance analysis, an approach taken, for instance, by  \cite{baratella2020} for very young dwarfs,  and/or more realistic model atmospheres accounting for chromospheric activities and magnetic fields.


 


\section{Summary and conclusions}

In this work, we have studied high-resolution, high-quality spectra for 40 red clump stars in 18 OCs and determined the abundance ratios of four $\alpha$ elements, O, Mg, Si, and Ca, as well as those of Li, Na, Al, and Ti.
Our sample does not contain any Li-rich giant. While the behavior of Li follows the overall expected depletion during the stellar ascent on the giant branch, abundances do show a clear scatter in Li abundance at any given $T_{\textrm{eff}}$ (or  $\log$ g) across the parameter range, hinting at other factors playing a role (e.g., age, metallicity and binarity). Given the small parameter space covered by our sample, no further deduction was attempted, and we defer to dedicated studies, such as those based on Gaia-ESO data for clusters and field stars \citep{magrini21a,magrini21b,romano21}.

We performed extensive comparison with the literature, both directly in terms of previously studied clusters and field stars (HARPS-GTO nearby dwarf sample and APOGEE DR17 field disk stars). Overall, our results are in good agreement with literature clusters (five clusters in common with the literature), and the only significant offset, even if small, is observed for Mg among younger clusters, while such an offset is not observed for the keystone cluster NGC 2682. The comparison with field stars is of limited value as both HARPS-GTO and APOGEE DR17 thin-disk-giant samples have an age distribution quite different from our cluster sample, with a very limited number of stars spanning the appropriate age range. However, we observe no major offsets when considering the youngest among the APOGEE DR17 thin-disk giants.

We considered the Na and Al abundances in our clusters to explore the behavior of these elements with respect to stellar evolution and nucleosynthesis. We observe an overall mild increasing trend of Na$_{NLTE}$ with respect to mass, in qualitative agreement with theoretical predictions, which are, however, calculated at metallicities that cover only the metal-poor end of the distribution of our clusters. Our data span a larger Na range than the predictions and seem to confirm an earlier suggestion from \cite{smiljanic16,smiljanic18}, based on a smaller sample, of the presence of some Na enhancement even among stars below 2\,M$_{\sun}$. 
We find no convincing trend for Al with stellar mass and the observations present a larger scatter than that spanned by models, especially concerning the most metal-rich stars among the clusters in our sample.

We investigated the Galactic gradients of abundances as a function of Galactocentric radius, distance from the Galactic midplane and age. In order to have a larger sample with a wider age distribution, we take into considerations also clusters from other recent studies (APOGEE, Gaia-ESO, GALAH and OCCASO) as well as clusters previously studied within the SPA program.
We find a significant Mg radial gradient both for present study (calculated for young clusters) and including all ages, in excellent agreement with recent literature findings.
For Ca we also find a significant radial gradient, a value between that of \cite{casamiquela19}, who reports a flat trend, and  that of \cite{donor20}.

Significant vertical positive gradients were found for Mg and Ca, in agreement with commonly reported findings of increasing $\alpha$-elements moving away from the thin disk.
Magnesium and Ca also show a significant age gradient, while all other elements are consistent with a flat distribution, in qualitative agreement with \cite{casamiquela19}.

Finally, we have compared the Galactocentric distribution of Mg and O in the combined sample with state-of-the-art chemodynamical models. For clusters older than $\sim$ 0.6 Gyr, the overall distribution is in very good agreement with results of the MCM models for both O and Mg.
For younger clusters the observations show a larger scatter and the agreement with the models becomes poorer for Mg and especially for O.

While there is naturally room for improvement in the chemo-dynamical models, it seems likely that at least part of this effect is due to the unsuitability of traditional 1D analysis and model atmospheres to derive atmospheric parameters and abundance ratios in very young giants, as effects of phenomena, such as chromospheric activity and magnetic fields, are not accounted for. 
These problems in young stars have essentially only been examined in the case of main-sequence or pre-main-sequence stars. Our findings show clearly for the first time that the effect(s) extend also to the derivation of Mg and O elemental abundances in giants. This is not a surprising result; in fact, the modeling of atmospheres and spectra for giants is even more challenging than for dwarfs.\\

\begin{acknowledgements}
We thank the TNG personnel for help during the observations and I. This research used the facilities of the Italian Center for Astronomical Archive (IA2) operated by INAF at the Astronomical Observatory of Trieste. This work exploits the Simbad, Vizier, and NASA-ADS
databases and the software TOPCAT \citep{topcat}.  This work has made use of data from the European Space Agency (ESA) mission Gaia (https://www.cosmos.esa.int/gaia), processed by
the Gaia Data Processing and Analysis Consortium (DPAC,
https://www.cosmos.esa.int/web/gaia/dpac/consortium). Funding
for the DPAC has been provided by national institutions, in particular the institutions participating in the Gaia Multilateral Agreement.
We acknowledge funding from MIUR Premiale 2016 MITiC. 
This work was partially funded by the PRIN INAF 2019 grant ObFu 1.05.01.85.14 (“Building up the halo: chemo-dynamical tagging in the age of large surveys”, PI. S. Lucatello) and by the German
      \emph{Deut\-sche For\-schungs\-ge\-mein\-schaft, DFG\/} project
      number Ts~17/2--1.
      GC acknowledges support from the European Research Council Consolidator Grant funding scheme (project ASTEROCHRONOMETRY, G.A. n. 772293, http://www.asterochronometry.eu). X.F. acknowledges the support of China Postdoctoral Science Foundation No. 2020M670023, the science research grants from the China Manned Space Project with NO.CMS-CSST-2021-A08, the National Key R\&D Program of China No. 2019YFA0405500, and the National Natural Science Foundation of China (NSFC) under grant No.11973001, 12090040, and 12090044.
\end{acknowledgements}

\clearpage

\begin{appendix}
\section{Results on individual stars}
\renewcommand{\thetable}{A\arabic{table}}
\onecolumn
\begin{table*}[ht]
\setlength{\tabcolsep}{1.25mm}
\begin{center}
\caption{Chemical abundances (O and Mg) of observed targets}
\begin{tabular}{lccccccccccc}
\label{tab:A1}\\
\hline\hline
 Name        &     Gaia ID &$T_{\textrm{eff}}$ &log g& [Fe/H]&v$_{micro}$& [O/Fe]&$\sigma_1$&N &[Mg/Fe]&$\sigma_1$&N\\
             &    & (K)&(dex)  & (dex) &km/s & (dex)  &(dex)  &   &(dex)  &(dex) & \\
\hline
ASCC 11  &241730418805573760&5250&2.15&-0.14&2.4&-0.179&0.05&1&0.055 &0.12 &4 \\
Alessi 1\_1 &402506369136008832&5000&2.65&-0.10&1.5&0.005 &0.15&2&-0.039&0.06&5 \\
Alessi 1\_2 &402505991178890752&5200&3.20&0.08&1.5 &0.125 &0.08&2&-0.195 &0.15&4 \\
Alessi 1\_3 & 	402867593065772288&5250 &3.27&0.07&1.6&0.080 &0.07&2&-0.184 &0.07&5 \\ 
Alessi 1\_4 & 402880684126058880&5120&3.09&-0.05&1.6 &0.121 &0.09&2&-0.115 &0.10&5 \\
Alessi-Teusch 11 &2184332753719499904&4560&2.10 &-0.19&2.1&0.143 &0.07&1&-0.170&0.06&4\\
Basel 11b\_1 &3424056131485038592&5180&3.15	 &0.00&2.2&0.108 &0.06&2&-0.140&0.06&5 \\
Basel 11b\_2 &3424055921028900736&5220&3.07 &0.02&2.2&0.150 &0.04&2&-0.184 &0.05&5 \\
Basel 11b\_3  &3424057540234289408&4950&2.83 &-0.04&2.1 &0.003 &0.13&2&-0.143&0.05&5 \\
COIN-Gaia 30 &532533682228608384&5200&3.40 &0.03&2.3 &0.307 &0.06&2&-0.184&0.06& 4\\
Collinder 463\_1 &534207555539397888&4730&2.12 &-0.20&2.5&0.080 &0.12&2&-0.172&0.05&5 \\
Collinder 463\_2 & 534363067715447680 &4730&2.30&-0.10&2.4&0.051 &0.06&2&-0.143&0.05&4 \\
Gulliver 18  & 1836389309820904064&4590&2.60 &-0.10&2.8&0.340 &0.08&2&-0.278&0.07&5  \\
Gulliver 24 &430035249779499264&4450&2.65  &-0.18&2.0 &0.258 &0.07&1&-0.169&0.04&4 \\
Gulliver 37  & 2024469226291472000&4850&3.65 &0.10&0.8 &0.046&0.07&2&-0.290&0.09&4 \\
NGC 2437\_1 &3029609393042459392 &5050&2.77&0.04&2.1 &0.007 &0.05&2&-0.213&0.05&4 \\
NGC 2437\_2 &3029202711180744832&5250&3.32 &0.07&2.2&0.171 &0.07&2&-0.235&0.07&3\\
NGC 2437\_3 & 3030364134752459904&5300&3.13 &-0.05&2.3&0.107 &0.05&2&-0.165 &0.08&5 \\
NGC 2437\_4   & 3029132686034894592&5085&2.90 &0.00&1.7&0.049 &0.07&2&-0.134&0.09 &4\\ 
NGC 2437\_5 &3029156222454419072&5030&2.85 &0.00&1.7 &0.026 &0.09&2&-0.168&0.10 &5 \\
NGC 2437\_6 &3029207006148017664&4990&2.72 &0.00&2.1 &-0.028 &0.05&2&-0.292&0.06&4 \\
NGC 2437\_7 & 3029226694277998080&4650&2.27 &-0.07&1.2&-0.085 &0.10&2&-0.219&0.08&4 \\
NGC 2509 & 5714209934411718784&4705&2.53  &-0.10&1.5&-0.028 &0.13&2&-0.038&0.09&3\\
NGC 2548\_1	& 3064481400744808704&5370&3.67 &0.00&2.0&0.289 &0.16&2&-0.190 &0.07&3 \\
NGC 2548\_2	&3064537647636773760&5050&2.65 &-0.02&1.6 &0.019 &0.16&2&-0.138 &0.05&5 \\
NGC 2548\_3    &3064579703955646976&4930&2.70&-0.01&1.6 &0.001 &0.07&2&-0.115 &0.07 &4 \\
NGC 2548\_4 	&3064486692144030336&5200&3.18 &-0.07&1.2 &0.084 &0.09&2&-0.184 &0.07 &3 \\
NGC 7082 	&1972288740859811072&4994&2.25 &-0.15&3.0 &-0.019 &0.07&2&-0.120&0.08&4  \\
NGC 7209\_1 	&1975004019170020736 &4880&2.35&0.00&1.7&-0.067 &0.09&2&-0.132 &0.06&5 \\
NGC 7209\_2 	 &1975002919658397568&4600&2.79 &-0.07&2.5 &0.421 &0.09&2&-0.149 &0.10&5 \\
Tombaugh 5\_1 	 & 473266779976916480&5010&3.17 &0.04&2.2 &0.110 &0.11&2&-0.148&0.05& 3 \\
Tombaugh 5\_2 	 & 473275782228263296&4900&2.31 &-0.07&2.0 &-0.040 &0.14&2&-0.024&0.07&3  \\
Tombaugh 5\_3 	 & 473266779976916480&5150&3.08 &0.07&2.2 &0.211 &0.07&2&-0.094 &0.07&4 \\
UPK 219    & 2209440823287736064&5203&3.01 &0.02&2.7 &0.076 &0.06&2&-0.174 &0.11&4 \\
\hline    
Collinder 350\_1   &4372743213795720704&4200&1.30	&-0.40&2.2&-0.003 &0.08&2&0.060 &0.07&5 \\
Collinder 350\_2  &4372572888274176768&5300&3.15 &0.02&1.6  &0.083 &0.10&2&-0.029 &0.11&5 \\ 
NGC 2682\_1    &604921512005266048&4687&2.37 	 &-0.05&1.5&0.089 &0.15&2&-0.001 &0.06&3 \\
NGC 2682\_2    &604920202039656064 &4900&2.76	 &0.02&1.7 &0.167 &0.14&2&0.008&0.08& 4\\
NGC 2682\_3  	&604904950611554432&5000&2.77 &-0.03&1.2  &0.036 &0.08&2&-0.049 &0.08&4\\
NGC 2682\_4  	&604917728138508160&5195&3.25&0.05&1.3 &0.084 &0.12&2&-0.057&0.07&4\\
SUN             &~~~&5770&4.44&0.03&1.0&0.082 &0.04&2 &0.011&0.09&5\\
\hline
\hline
\end{tabular}
\end{center}
\end{table*}
\twocolumn

\onecolumn
\begin{table*}[ht]
\setlength{\tabcolsep}{1.25mm}
\begin{center}
\caption{Chemical abundances (Si, Ca, and Ti) of observed targets. }
\label{tab:A2}
\begin{tabular}{lccccccccc}
\hline\hline
  Name        &[Si/Fe]&$\sigma_1$&N&[Ca/Fe]&$\sigma_1$&N&[Ti/Fe]&$\sigma_1$&N\\
             &(dex) &(dex)  &  &(dex) & (dex) & &(dex)   &(dex)&  \\
\hline
ASCC 11  &0.174 & 0.07&3 &0.129 & 0.15&12&-0.171&0.12&70\\
Alessi 1\_1 &0.184 &0.05& 9&0.027 &0.10&26&-0.132& 0.11&70 \\
Alessi 1\_2 &0.112 &0.12&9&-0.054 &0.14&24&-0.101&0.15& 70\\
Alessi 1\_3&0.087 &0.05&9&-0.049 &0.09&24&-0.028&0.11&70 \\ 
Alessi 1\_4 &0.184 &0.17&9&-0.026 &0.12&25&-0.135&0.14&65 \\
Alessi-Teusch 11 &0.267 &0.07&9&-0.143 &0.13&25&-0.132&0.13&60\\
Basel 11b\_1&0.064 &0.04&6&-0.063 &0.13&24&-0.020&0.07 &50\\
Basel 11b\_2&0.124 &0.04&7&-0.108 &0.09&23&-0.040&0.08&30 \\
Basel 11b\_3  &0.092 &0.04&7&-0.076 &0.09&25&-0.070&0.09&50 \\
COIN-Gaia 30 &0.000&0.04&9&-0.151 &0.09&25&0.093&0.10&50\\
Collinder 463\_1 &0.141 &0.05&8&-0.179 &0.07&21 &-0.074&0.08&30\\
Collinder 463\_2 &0.171 &0.04&8&-0.173 &0.08 &23&-0.173&0.07&30 \\
Gulliver 18   &0.164&0.09&6&-0.176 &0.17&22&0.227&0.22&100 \\
Gulliver 24  &0.288 &0.08&6&-0.123 &0.12&22&0.097&0.15&50 \\
Gulliver 37  &0.077 &0.07&7&-0.214&0.19&15&-0.199& 0.27&50\\
NGC 2437\_1 &0.080&0.04&9&-0.128 &0.08&22&-0.066&0.09&50 \\
NGC 2437\_2&0.113 &0.05&6&-0.097 &0.11&22&-0.084&0.13&75 \\
NGC 2437\_3&0.104 &0.06&7&-0.079 &0.12&21&-0.023&0.16&110 \\
NGC 2437\_4   &0.126 &0.06&9&-0.022 &0.15&25&-0.063&0.14&65\\ 
NGC 2437\_5 &0.167 &0.08&8&-0.104 &0.14&25&-0.160&0.15&75 \\
NGC 2437\_6 &0.078 &0.04&6&-0.146 &0.10 &22&-0.172&0.11&65 \\
NGC 2437\_7 &0.170 &0.07&7&-0.153 &0.14&22&-0.160&0.19&90 \\
NGC 2509 &0.269 &0.10&7&0.024 &0.17&24&-0.020&0.12& 50\\
NGC 2548\_1	&0.062 &0.04&7&-0.025 &0.11&23&0.010&0.10& 40\\
NGC 2548\_2	&0.057 &0.04&8&-0.044 &0.11&25&-0.097&0.13& 70\\
NGC 2548\_3    &0.099&0.05&9&0.025 &0.11&26&-0.073&0.09& 50\\
NGC 2548\_4 	&-0.017 &0.04&9&-0.123 &0.08&20&-0.195&0.09&50\\
NGC 7082 &0.085 &0.05&6&0.021 &0.10&20&-0.041&0.10 &50  \\
NGC 7209\_1  &0.103&0.05&9&-0.102 &0.12&24&-0.086&0.12&50 \\
NGC 7209\_2 	  &0.186&0.09&7&-0.07 &0.18&22&0.045&0.12&30 \\
Tombaugh 5\_1 	 &0.086&0.07&7&-0.171 &0.10&21&-0.078&0.10&50 \\
Tombaugh 5\_2 	  &0.137&0.04&6&-0.038&0.09&21 &0.053&0.12& 50\\
Tombaugh 5\_3 	 &0.121&0.05&7&-0.095 &0.11&23&0.075&0.11&50 \\
UPK 219     &0.073 &0.05&8&-0.167 &0.13&22&0.059&0.16&150 \\
\hline    
Collinder 350\_1   &0.125 &0.07&8&0.009 &0.12&23&0.024&0.13&50 \\
Collinder 350\_2   &-0.008&0.05&6&-0.024 &0.10&23&-0.013&0.09&50 \\
NGC 2682\_1   &0.098&0.06&6&-0.086 &0.14&23&-0.067&0.15&50 \\
NGC 2682\_2    &0.094&0.03&8&-0.144 &0.08& 23&-0.097&0.19&110 \\
NGC 2682\_3  & 0.091&0.04&8&-0.044 &0.08&24&-0.070&0.12&50 \\
NGC 2682\_4  	&0.056 &0.05&8&0.031 &0.09&21 &0.030&0.12&50 \\
SUN            &0.05 &0.03 &9&-0.053&0.05 &25&-0.033&0.06&130\\
\hline\hline
\end{tabular}
\end{center}
\end{table*}
\twocolumn

\begin{table*}[ht]
\setlength{\tabcolsep}{1.25mm}
\begin{center}
\caption{Lithium, sodium and aluminum abundances with  NLTE correction.}
\begin{tabular}{lrrrrrrr}
\hline\hline
  Name  & log $\epsilon$(Li)    &$\sigma_{log (Li)}$ & [Na/Fe]&$\sigma_{[Na/Fe]}$   &[Al/Fe]&$\sigma_{[Al/Fe]}$ \\
  &(dex)& (dex)&(dex)  &(dex)&(dex)&(dex)\\
\hline
ASCC 11     & 1.44    &0.09 & 0.50&0.08  &0.18& 0.07   \\
Alessi 1\_1   &0.71    & 0.09 &0.09&0.06 &-0.17&0.06   \\
Alessi 1\_2  &0.97   &0.10 & -0.12&0.10 &-0.06&0.11  \\
Alessi 1\_3  & 0.60 &0.14 &-0.03&0.06 &-0.22&0.07  \\
Alessi 1\_4  &0.83    & 0.03&0.00 &0.10&-0.11&0.11 \\
Alessi-Teusch 11 &0.53     &0.26&0.06 &0.06  &-0.07&0.06  \\ 
Basel 11b\_1 	 &1.47  &0.11 &0.05&0.05 &-0.02&0.06 \\ 
Basel 11b\_2     &1.30  & 0.08 &0.07 &0.03 & -0.16&0.04 \\
Basel 11b\_3 &0.38     &0.28&0.07&0.05 &-0.07&0.05  \\
COIN-Gaia 30 &1.70   &0.26&0.13&0.09 & -0.11 &0.06\\ 
Collinder 463\_1 &0.58     &0.06&0.17&0.05  &-0.11&0.05 \\
Collinder 463\_2 &0.63    &0.07&0.19&0.05 &-0.08 &0.05\\
Gulliver 18 &0.80    &0.12 & 0.31&0.11 &0.12 &0.09 \\
Gulliver 24	&-0.26    &  upper limit& 0.03&0.05&0.02&0.06  \\
Gulliver 37	&1.42     &0.07&-0.15&0.05 &-0.13&0.09  \\
NGC 2437\_1 &1.45    &0.10&0.08 &0.06 &-0.05&0.06 \\ 
NGC 2437\_2 &1.49    &0.18 &-0.04 &0.06&0.07&0.06  \\
NGC 2437\_3 &1.55     &0.28&0.11 &0.07 &0.08&0.08\\
NGC 2437\_4 &1.35 &0.07&0.11 &0.07 &-0.08&0.06 \\
NGC 2437\_5 &1.11    &0.04&0.08 &0.10 &-0.17&0.11\\
NGC 2437\_6 &1.32    &0.07 &0.09 &0.06& 0.02&0.06\\
NGC 2437\_7 &0.28     &0.18&0.00 &0.11 &-0.10&0.10 \\
NGC 2509    &-0.02 &   upper limit&0.21 &0.07&0.05&0.09\\ 
NGC 2548\_1 &1.59   &0.08&0.07 &0.05 &-0.19&0.08  \\
NGC 2548\_2 & 0.69    &0.18 &0.01 &0.05 &-0.12&0.04\\
NGC 2548\_3 &1.02    &0.05&0.12 &0.05 & -0.06&0.06\\
NGC 2548\_4 &1.29   &0.08&-0.13 &0.06  & -0.17&0.07 \\
NGC 7082 &1.52    & 0.08&0.23 &0.06 & 0.04&0.06 \\ 
NGC 7209\_1 &0.45    & 0.19&-0.07 &0.05 &  -0.11&0.05 \\
NGC 7209\_2 &1.27     &0.16&0.18 &0.05 & 0.12&0.06 \\
Tombaugh 5\_1 	&1.06     &0.18&0.22  &0.05  &-0.12&0.06 \\
Tombaugh 5\_2 &0.82 &0.13&0.20 &0.06 & -0.05&0.06 \\
Tombaugh 5\_3 &1.01    & 0.04& 0.31 &0.06& -0.14&0.06\\
UPK 219 &1.10     &0.13 &0.32 &0.09 & -0.02&0.10  \\
\multicolumn{7}{c}{Comparison clusters} \\
Collinder 350\_1 &0.37&0.07 &0.12&0.09& 0.18&0.09\\
Collinder 350\_2 &1.58     &0.14 &0.19 &0.05 &-0.17&0.06 \\
NGC 2682\_1 &-0.30     &upper limit &0.00&0.06&0.00&0.07\\ 
NGC 2682\_2 &0.56     & 0.27&0.09&0.09 & -0.13&0.11\\
NGC 2682\_3 &0.58     &upper limit &0.06&0.05&-0.04&0.05 \\
NGC 2682\_4 &0.57     &upper limit&-0.07&0.14 &-0.02&0.04\\
SUN &0.96 &0.01 &-0.03&0.07&-0.03&0.07\\
\hline
\hline
\end{tabular}
\label{tab:Li}
\end{center}
\end{table*}

\begin{table*}[ht]
\setlength{\tabcolsep}{1.25mm}
\begin{center}
\caption{Sensitivity matrix of Li, Na, and Al. 
}
\begin{tabular}{lrrrr}
\hline\hline
 Name   &$\Delta$log $\epsilon$(Li)/$\Delta$ T$_{\textrm{eff}}$  & $\Delta$log $\epsilon$(Li)/$\Delta \log$ g  &  $\Delta$log $\epsilon$(Li)/$\Delta$[Fe/H] & $\Delta$log $\epsilon$(Li)/$\Delta$ v$_{micro}$\\
\hline
Collinder 350\_1 &0.028&0.013&0.020&-0.003\\
Alessi-Teausch 11 &-0.026&0.030&0.030&-0.023 \\
Collinder 463\_1 &0.090&0.020&-0.007&0.001\\
Gulliver 37 &0.068&0.030&-0.003&0.004\\
NGC 2548\_3 &0.079&0.021&-0.043&0.000\\
Tombaugh 5\_1 &0.091&-0.002&0.001&0.000\\
Alessi 1\_4 &0.036&0.014&-0.012&0.013\\
UPK 219 &0.072&-0.031&-0.086&-0.069\\
NGC 2548\_1 &0.060&-0.001&0.030&0.049\\
\hline\hline
  Name   &$\Delta$[Na/Fe]/$\Delta$ T$_{\textrm{eff}}$  & $\Delta$[Na/Fe]/$\Delta$ $\log$ g  &  $\Delta$[Na/Fe]/$\Delta$[Fe/H] & $\Delta$[Na/Fe]/$\Delta$ v$_{micro}$\\
\hline
Collinder 350\_1&0.011&0.073&-0.049&-0.004\\
Alessi Teusch 11&0.037&-0.013&-0.047&-0.028\\
Collinder 463\_1&0.021&-0.009&-0.047&0.002\\
Gulliver 37&0.034&-0.015&-0.026&-0.028\\
NGC2548\_3&0.031&-0.008&-0.036&-0.017\\
Tombaugh 5\_1&0.032&-0.020&-0.043&0.001\\
Alessi 1\_4&0.000&0.042&-0.092&-0.012\\
UPK219&0.084&-0.012&-0.047&0.000\\
NGC2548\_1&0.021&-0.011&-0.044&0.000\\
\hline\hline
  Name   &$\Delta$[Al/Fe]/$\Delta$ T$_{\textrm{eff}}$  & $\Delta$[Al/Fe]/$\Delta$ $\log$ g  &  $\Delta$[Al/Fe]/$\Delta$[Fe/H] & $\Delta$[Al/Fe]/$\Delta$ v$_{micro}$\\
\hline
Collinder 350\_1&0.027&-0.002&-0.053&-0.087\\
Alessi Teusch 11&0.036&0.007&-0.051&-0.027\\
Collinder 463\_1&0.021&0.002&-0.050&-0.008\\
Gulliver 37&0.031&-0.007&-0.029&-0.09\\
NGC2548\_3&0.064&0.000&-0.052&-0.012\\
Tombaugh 5\_1&0.030&0.000&-0.051&-0.011\\
Alessi 1\_4&0.045&0.000&-0.051&-0.011\\
UPK219&0.093&0.000&-0.050&-0.015\\
NGC2548\_1&0.064&0.000&-0.052&-0.012\\
\hline
\end{tabular}
\label{tab:sens2}
\end{center}
\end{table*}

\begin{table*}[ht]
\setlength{\tabcolsep}{1.25mm}
\begin{center}
\caption{Sensitivity matrix of $\alpha$-elements.
}
\begin{tabular}{lrrrr}
\hline\hline
  Name   &$\Delta$[O/Fe]/$\Delta$ T$_{\textrm{eff}}$  & $\Delta$[O/Fe]/$\Delta$ $\log$ g  &  $\Delta$[O/Fe]/$\Delta$[Fe/H] & $\Delta$[O/Fe]/$\Delta$ v$_{micro}$\\
\hline
Collinder 350\_1&0.011&0.073&-0.049&-0.004\\
Alessi Teusch 11&0.007&0.065&-0.030&-0.002\\
Collinder 463\_1&0.004&0.044&-0.030&-0.001\\
Gulliver 37&0.009&0.048&-0.018&-0.004\\
NGC2548\_3&0.006&0.042&-0.023&-0.001\\
Tombaugh 5\_1&0.006&0.063&-0.029&0.000\\
Alessi 1\_4&0.006&0.024&-0.064&-0.003\\
UPK219&0.023&0.046&-0.030&-0.001\\
NGC2548\_1&0.005&0.085&-0.030&-0.002\\
\hline\hline
  Name   &$\Delta$[Mg/Fe]/$\Delta$ T$_{\textrm{eff}}$  & $\Delta$[Mg/Fe]/$\Delta$ $\log$ g  &  $\Delta$[Mg/Fe]/$\Delta$[Fe/H] & $\Delta$[ Mg/Fe]/$\Delta$ v$_{micro}$\\
\hline
Collinder 350\_1 &0.019&-0.010&-0.068&-0.031\\
Alessi Teusch\_11 &0.025&-0.003&-0.045&-0.040\\
Collinder 463\_1 &0.019&-0.004&-0.046&-0.020\\
Gulliver 37 &0.017&-0.011&-0.023&-0.085\\
NGC2548\_3 &0.030&-0.010&-0.045&0.054\\
Tombaugh 5\_1 &0.029&-0.015&-0.045&-0.021\\
Alessi 1\_4 &0.036&-0.008&-0.092&-0.017\\
UPK219 &0.094&-0.012&-0.049&-0.034\\
NGC2548\_1 &0.039&-0.023&-0.047&-0.027\\
\hline\hline
  Name   &$\Delta$[Si/Fe]/$\Delta$ T$_{\textrm{eff}}$  & $\Delta$[Si/Fe]/$\Delta$ $\log$ g  &  $\Delta$[Si/Fe]/$\Delta$[Fe/H] & $\Delta$[Si/Fe]/$\Delta$ v$_{micro}$\\
\hline
Collinder 350\_1&-0.030&0.030&-0.058&-0.012\\
Alessi Teusch 11&-0.020&0.048&-0.043&-0.049\\
Collinder 463\_1&-0.001&0.025&-0.042&-0.011\\
Gulliver 37&-0.020&0.022&-0.020&-0.087\\
NGC2548\_3&-0.001&0.024&-0.037&-0.051\\
Tombaugh 5\_1&-0.006&0.040&-0.059&-0.020\\
Alessi 1\_4&0.003&0.011&-0.082&-0.018\\
UPK219&0.022&0.020&-0.043&-0.033\\
NGC2548\_1&0.013&0.025&-0.035&-0.017\\
\hline\hline
  Name   &$\Delta$[Ca/Fe]/$\Delta$ T$_{\textrm{eff}}$  & $\Delta$[Ca/Fe]/$\Delta$ $\log$ g  &  $\Delta$[Ca/Fe]/$\Delta$[Fe/H] & $\Delta$[Ca/Fe]/$\Delta$ v$_{micro}$\\
\hline
Collinder 350\_1&0.060&-0.010&-0.082&-0.058\\
Alessi Teusch 11&0.000&-0.002&-0.216&0.000\\
Collinder 463\_1&0.030&-0.003&-0.055&-0.039\\
Gulliver 37&0.047&-0.034&-0.026&-0.186\\
NGC2548\_3&0.054&-0.013&-0.050&-0.049\\
Tombaugh 5\_1&0.051&-0.019&-0.051&-0.037\\
Alessi 1\_4&0.057&-0.087&-0.106&-0.040\\
UPK219&0.134&-0.008&-0.054&-0.069\\
NGC2548\_1&0.056&-0.027&-0.050&-0.062\\
\hline\hline
  Name   &$\Delta$[Ti/Fe]/$\Delta$ T$_{\textrm{eff}}$  & $\Delta$[Ti/Fe]/$\Delta$ $\log$ g  &  $\Delta$[Ti/Fe]/$\Delta$[Fe/H] & $\Delta$[Ti/Fe]/$\Delta$ v$_{micro}$\\
\hline
Collinder 350\_1&0.087&0.010&-0.078&-0.066\\
Alessi Teusch 11&0.077&0.000&-0.051&-0.100\\
Collinder 463\_1&0.045&0.000&-0.051&-0.023\\
Gulliver 37&0.064&-0.007&-0.029&-0.262\\
NGC2548\_3&0.067&-0.002&-0.042&-0.051\\
Tombaugh 5\_1&0.066&-0.004&-0.051&-0.039\\
Alessi 1\_4&0.082&-0.002&-0.106&-0.034\\
UPK219&0.201&-0.006&-0.0555&-0.038\\
NGC2548\_1&0.048&-0.008&-0.055&-0.075\\
\hline
\end{tabular}
\label{tab:sens1}
\end{center}
\end{table*}

\clearpage
\small
\onecolumn
\begin{landscape}
\begin{longtable}{lcccccccccccccccccc}
\caption{Comparison of chemical abundance with literature.
}
\label{tab-comp}\\
\hline\hline
  star &[Mg/Fe]&  $\sigma$[Mg/Fe] &[Si/Fe]&  $\sigma$[Si/Fe]   &    [Ca/Fe] & $\sigma$[Ca/Fe] &   [O/Fe] & $\sigma$[O/Fe] & [Ti/Fe] & $\sigma$[Ti/Fe] & [Na/Fe] & $\sigma$[Na/Fe]&[Al/Fe]&$\sigma$[Al/Fe] &Ref (NLTE)\\
        Gaia ID         &   (K) & &(dex)& &(dex)& &(dex) &&(dex)&&(dex)&&(dex)\\
\hline
NGC 2682\_1 &-0.001&0.060&0.098&0.060&-0.086&0.140&0.089&0.150&-0.067&0.150 &0.002&0.060&0.000&0.070& Present study  \\
             &0.130&0.280&0.270&0.260&-0.080&0.340&-0.080 &0.070 &-0.130&0.210&0.100&0.260&0.370&0.350 &(1)(NO)\\
             &-0.011&0.009&0.055&0.009&0.003&0.009&0.001&0.010&0.005&0.010&0.100&0.030& 0.066&0.016& (2)(NO) \\
             &-0.023&0.014 &0.044&0.034 &-0.016&0.045 &-0.049&0.069 &0.036&0.034 &&&&&(3)  \\
             &0.010&0.040&0.060&0.040&&&-0.040&0.150&&&0.140&0.050    &0.100 &0.040&(4)(YES)\\
             604921512005266048          &   & && && & &&&&\\

\hline
NGC 2682\_2 	 &0.008&0.080&0.094&0.030&-0.144&0.080&0.167&0.140&-0.097&0.190 &0.087&0.090&-0.130&0.110& Present study   \\
             &0.220&0.270&0.160&0.200&-0.120&0.220&&&-0.150&0.200&0.010&0.200 &0.300&0.420     &(1) (NO) \\
             &-0.009&0.009&0.003&0.009&0.002&0.010&-0.070&0.010&-0.050&0.010&0.130&0.030  &-0.009&0.017&(2)(NO)  \\
             &-0.002&0.027 &0.052&0.033 &0.022&0.027 &0.115&0.069 &0.044&0.033 & &   &&&(3)\\
             &-0.120&0.040&-0.030&0.050&&&-0.060&0.140&&&0.040   &0.050&0.130&0.030&(4) (YES)\\
             604920202039656064           &   & && && & &&&&\\
\hline
NGC 2682\_3  &-0.049&0.080&0.091&0.030&-0.044&0.080&0.036&0.080&-0.070&0.120&0.061&0.050 &-0.040&0.050& Present study  \\
            &0.280&0.270&0.190&0.210&-0.050&0.220&&& 0.000&0.180&0.100&0.250 &0.390&0.280  &(1) (NO)\\ 
            &-0.020&0.010&-0.020&0.010&-0.020&0.010&&&0.010&0.010&0.080&0.040 &-0.014&0.018&(2)(NO)  \\
            &0.080&&0.090&&0.080&&&&0.050&&0.270&    &0.160&&(5)(NO)\\
            &0.003&0.032&0.114&0.027&0.011&0.029&0.068&0.054&0.037&0.018&0.169&0.023&0.159&0.024&(6)(NO)\\
             604904950611554432          &   & && && & &&&&\\
\hline
NGC 2682\_4 	  &-0.057&0.070&0.056&0.050&0.031&0.090 &0.084&0.120&0.030&0.120&-0.068&0.140&-0.020&0.040&Present study \\
              &0.270&0.280&0.200&0.220&-0.070&0.340& & &-0.040&0.240&0.120&0.230 &0.290&0.310    &(1)(NO)\\
              &-0.04&0.010&0.005&0.010&0.050 & 0.010&&&-0.080&0.020&-0.080&0.040 &-0.025&0.018&(2)(NO)\\
              &0.068&0.032 &0.280&0.027&0.103&0.031&0.264&0.057&-0.027&0.023&0.091 & 0.023 &0.277&0.026&(6) (NO)\\
              604917728138508160         &   & && && & &&&&\\
\hline
\hline
Collinder 350\_2  &-0.029&0.110&-0.008&0.050&-0.024&0.100&0.083&0.100&-0.013&0.090&0.190&0.050 &-0.170&0.060& Present study&\\
                 &0.050&0.150&0.030&0.120&-0.010&0.120&& &0.000&0.140&0.250&0.100& -0.010&0.100&(7) (NO)\\
                 4372572888274176768           &   & && && & &&&&\\
\hline
Basel 11b\_3      &-0.143 &0.050 &0.092&0.040 &-0.076&0.090&0.003&0.130&-0.070&0.090&0.069&0.050 &-0.060&0.050&Present study&\\
                 &-0.040 &0.009 &0.006 &0.009 &-0.015 &0.010&-0.02&0.010&-0.050&0.010&0.040&0.030&-0.071&0.017&(2)(NO)\\
                 3424057540234289408          &   & && && & &&&&\\
\hline
Tombaugh 5\_1     &-0.148&0.050&0.086&0.100&-0.171&0.100&0.110&0.110&-0.078&0.100&0.222&0.050 &&& Present study\\
                 && &0.040&0.102 &-0.070&0.08&&&-0.090&0.132& &&&&(8)\\
                 473266779976916480          &   & && && & &&&&\\
\hline
NGC 2548\_1      &-0.190&0.070&0.062&0.040&-0.025&0.110&0.289&0.160&0.010&0.100&0.065&0.050 &-0.190&0.080&Present study\\
                &-0.067&0.033&-0.026&0.028&0.299&0.033&0.054 &0.061&-0.071&0.020&0.205&0.024&0.045&0.028&(6) (NO)\\
                 3064481400744808704           &   & && && & &&&&\\
\hline
NGC 2548\_2      &-0.138&0.050&0.057&0.040&-0.044&0.110&0.019&0.160&-0.097&0.130 &0.012&0.050&-0.120&0.040& Present study\\
                &-0.047&0.033&-0.012&0.027&0.260&0.031&-0.121&0.058&0.018&0.018&0.186&0.023 &0.035 &0.026&(6) (NO)\\
               3064537647636773760         &   & && && & &&&&\\
\hline
\end{longtable}
\tablefoot{The last column contains the reference from difference source and information of NLTE correction for Na and Al:(1) \cite{jacobson11}; (2) APOGEE DR16; (3) \citep{casamiquela19};(4) \cite{gao18}; (5) \cite{luck15};(6) \cite{spina21};(7) \cite{casali20} the result based on FAMA code;(8) \cite{baratella18} }

\end{landscape}

\end{appendix}
\end{document}